\documentclass[lettersize,journal]{IEEEtran}
\usepackage{amsmath,amsfonts}
\usepackage{amssymb}
\usepackage{algorithmic}
\usepackage{algorithm}

\usepackage{array}
\usepackage{textcomp}
\usepackage{stfloats}
\usepackage{url}
\usepackage{xcolor}
\usepackage{verbatim}
\usepackage{graphicx,color}
\usepackage{cite}
\usepackage{bm}
\usepackage{ microtype, hyperref}
\usepackage[export]{adjustbox}
\newcommand{\be}{\begin{equation}}
\newcommand{\ee}{\end{equation}}
\newcommand{\ba}{\begin{array}}
\newcommand{\ea}{\end{array}}
\newcommand{\bea}{\begin{eqnarray}}
\newcommand{\eea}{\end{eqnarray}}

\newcommand{\diag}{{\operatorname{diag}}}

\newcommand{\herm}{^H}  % Hermitian transpose
\newcommand{\tran}{^{ \emph{T}}}  % transpose

\usepackage[printonlyused]{acronym}
\acrodef{SIM}{stacked intelligent metasurface}
\acrodef{RIS}{reconfigurable intelligent surface}
\acrodef{MISO}{multiple-input single-output}
\acrodef{CSI}{channel state information}
\acrodef{SINR}{signal-to-interference-plus-noise ratio}
\acrodef{GP}{geometric programming}
\acrodef{GDA}{gradient descent-ascent}
\acrodef{KKT}{Karush-Kuhn-Tucker}
\acrodef{SOCP}{second-order cone programming}
\acrodef{SDR}{semi-definite relaxation}
\acrodef{RF}{radio frequency}
\acrodef{BS}{base station}
\acrodef{UE}{user equipment}
\acrodef{MIMO}{multiple-input multiple-output}
\acrodef{SNR}{signal-to-noise ratio}
\acrodef{CDF}{cumulative distribution function}
\acrodef{PDF}{probability density function}
\acrodef{1G}{first-generation}
\acrodef{5G}{fifth-generation}
\acrodef{6G}{sixth-generation}

\title{Max-Min Fairness for Stacked Intelligent Metasurface-Assisted Multi-User MISO Systems}
\author{Nipuni  Ginige \IEEEmembership{(Graduate Student Member, IEEE)}, Prathapasinghe Dharmawansa, \\ Arthur Sousa de Sena \IEEEmembership{(Member, IEEE)}, Nurul Huda Mahmood \IEEEmembership{(Member, IEEE)}, \\ Nandana Rajatheva \IEEEmembership{(Senior Member, IEEE)} and Matti Latva-aho \IEEEmembership{(Fellow, IEEE)} 
\thanks{Nipuni Ginige, Prathapasinghe Dharmawansa, Arthur S. de Sena, Nurul H. Mahmood, Nandana Rajateva, and Matti Latva-aho are with the Center for Wireless Communications, University of Oulu, 90570 Oulu, Finland (e-mail: nipuni.ginige@oulu.fi, pkaluwad24@univ.yo.oulu.fi,  arthur.sena@oulu.fi, nurulhuda.mahmood@oulu.fi, nandana.rajatheva@oulu.fi, matti.latva-aho@oulu.fi).}
\thanks{This work was supported by 6G Flagship (Grant Number 369116) funded by the Research Council of Finland.}}
\begin{document}
\maketitle
 \begin{abstract}
\Ac{SIM} is an emerging technology that uses multiple reconfigurable surface layers to enable flexible wave-based beamforming.
In this paper, we focus on an \ac{SIM}-assisted multi-user multiple-input single-output system, where it is essential to ensure that all users receive a fair and reliable service level. To this end, we develop two max-min fairness algorithms based on instantaneous channel state information (CSI) and statistical CSI. For the instantaneous CSI case, we propose an alternating optimization algorithm that jointly optimizes power allocation using geometric programming and wave-based beamforming coefficients using the gradient descent-ascent method. For the statistical CSI case, since deriving an exact expression for the average minimum achievable rate is analytically intractable, we derive a tight upper bound and thereby formulate a stochastic optimization problem. This problem is then solved, capitalizing on an alternating approach combining geometric programming and gradient descent algorithms, to obtain the optimal policies. Our numerical results show significant improvements in the minimum achievable rate compared to the benchmark schemes. In particular, for the instantaneous CSI scenario, the individual impact of the optimal wave-based beamforming is significantly higher than that of the power allocation strategy. Moreover, the proposed upper bound is shown to be tight in the low signal-to-noise ratio regime under the statistical CSI.
 \end{abstract}

\section{INTRODUCTION}
%%%%%%%%%%%%%%%%%%%%%%%%%%%%%%%%%%%%%%%%%%%%%%%%%%%%%%%%%%%%%%%%%%%%%%%
\IEEEPARstart{S}{tacked} intelligent metasurface (SIM) is a novel technology that uses multiple (refractive) \ac{RIS} layers to enable wave-domain beamforming \cite{SIM_intro,  SIM_intro2, SIM_BD_RIS, SIM_intro3, direnzo2024stateartstackedintelligent}. By manipulating electromagnetic waves at the analog level, \ac{SIM} eliminates the need for conventional digital precoding, thus reducing processing delays \cite{SIM_intro, SIM_BD_RIS}. Moreover, \ac{SIM}-based transceivers require fewer \ac{RF} chains, leading to lower hardware costs and energy consumption compared to traditional massive \ac{MIMO} systems \cite{SIM_intro2, SIM_BD_RIS}.

The transmit beamforming waveform of \ac{SIM}-aided systems needs to be optimised to reap the full advantages of its benefits. Mathematical optimization has been essential in wireless communication systems from \ac{1G} to \ac{5G} and will continue to be crucial in the upcoming \ac{6G} \cite{optimization_survey}.  Usually, optimization involves tuning the transmission parameters, such as the transmit power, transmit beamformers, \ac{RIS} phase shifts, etc., with the aim of maximizing (or minimizing) one or more objective functions like the sum rate, system throughput, or spectral efficiency while effectively managing interference \cite{optimization_wireless_intro,optimization_wireless_intro2,power_opt_GP,BD_RIS_optimization}.
%It plays a key role in resource allocation, maximizing sum rate, system throughput, or spectral efficiency while effectively managing interference \cite{optimization_wireless_intro,optimization_wireless_intro2,power_opt_GP}.
However, maximizing a given performance measure may result in unfair allocation where users with favorable conditions are prioritized over users in disadvantageous situations, such as cell edge users. 
In particular, max-min fairness is a widely used strategy to allocate resources efficiently, prioritizing weaker users without unnecessary waste \cite{max_min_def}.

Developing specialized schemes to enhance fairness by maximizing the minimum rate or \ac{SINR} is essential for \ac{SIM}-assisted networks, as their multilayer architecture presents unique optimization challenges that demand dedicated solutions \cite{max_min_intro2,max_min_intro}. In this regard, mathematical optimization is key, facilitating the joint optimization of power allocation and wave-based beamforming to fully harness the benefits of \ac{SIM} technology. Usually, resource optimization requires instantaneous \ac{CSI}, which incurs signaling overhead and practical limitations. Therefore, leveraging statistical \ac{CSI} reduces computational and signaling overhead compared to instantaneous \ac{CSI}, making \ac{SIM} optimization more practical for real-world applications \cite{SIM_optimization_statistical_CSI}. Although recent studies have focused on optimization techniques for \ac{SIM}-assisted networks targeting different objectives, ensuring fair resource allocation among users has remained largely unaddressed. Moreover, to the best of our limited knowledge, fairness-based resource optimization under statistical CSI constraints has not been addressed in the literature either.  
The next section reviews relevant related works.
%
%Max-min fairness is a well-known approach to allocating as much as possible to weaker users without unnecessarily wasting resources \cite{max_min_def}.  Thus, maximizing the wireless network's minimum rate or signal-to-interference-plus-noise ratio (SINR) is a suitable method to guarantee fairness in the network \cite{max_min_intro2,max_min_intro}. When we focus on \ac{SIM}-assisted wireless networks, we see that mathematical optimization is essential in performance improvement. Furthermore, optimization using statistical channel state information (CSI) reduces the computational overhead compared to instantaneous \ac{CSI} \cite{SIM_optimization_statistical_CSI}. 
%%%%%%%%%%%%%%%%%%%%%%%%%%%%%%%%%%%%%%%%%%%%%%
\subsection{RELATED WORKS}
Ample recent works on \ac{SIM}-assisted wireless networks have focused on power allocation and wave-based beamforming to maximize achievable sum rate or spectral efficiency \cite{an2023stackedintelligentmetasurfacesmultiuser, SIM_optimization_holographic_MIMO, SIM_optimization_statistical_CSI, SIM_fully_analog_beamforming, SIM_optimization_DRL,SIM_beamforming_DRL, hybrid_SIM}. 
The authors of \cite{an2023stackedintelligentmetasurfacesmultiuser} proposed a \ac{SIM}-based transceiver design for a \ac{MISO} downlink system. They formulated an optimization problem to iteratively determine the optimal power allocation and wave-based beamforming to maximize the system’s sum rate. An iterative water-filling algorithm was used for power allocation, while a projected gradient ascent algorithm and a successive refinement method were applied to address the beamforming problem. In \cite{SIM_optimization_holographic_MIMO}, an iterative projected gradient approach was applied to jointly optimize the transmit covariance matrix and \ac{RIS} phase shifts, which maximized the achievable rate in a multi-stream holographic \ac{MIMO} system with \ac{SIM}. Similarly, \cite{SIM_fully_analog_beamforming} proposed a \ac{SIM}-enhanced fully analog wideband beamforming design that approximated the end-to-end channel as a diagonal matrix by optimizing \ac{RIS} phase shifts. A deep reinforcement learning–based method was introduced in \cite{SIM_optimization_DRL, SIM_beamforming_DRL} to jointly optimize transmit power allocation and wave-based beamforming in a \ac{SIM}-assisted multi-user \ac{MIMO} system.  The authors of \cite{hybrid_SIM} proposed an optimal resource allocation algorithm while maximizing the sum rate for \ac{SIM}-assisted systems with both phase control and amplitude control layers. In \cite{shi2024harnessing}, a \ac{SIM}-enhanced cell-free massive \ac{MIMO} system was presented along with a greedy-based pilot allocation algorithm, an iterative optimization algorithm for wave-based beamforming, and a max-min spectral efficiency power control algorithm using a bisection method. Moreover, \cite{SIM_ISAC} proposed an algorithm that optimized wave-based beamforming by maximizing the sum rate and shaping the sensing beam pattern with a dual-normalized differential gradient descent.

Max-min fairness optimization was also extensively studied for conventional \ac{RIS}-assisted systems. In \cite{max_min_group_RIS}, max-min fairness in a group-transmissive \ac{RIS}-assisted downlink system was addressed using successive convex approximation and \ac{SDR}. The work in \cite{max_min_RIS} maximized the minimum weighted \ac{SINR}—subject to individual transmit power and \ac{RIS} reflection constraints—using \ac{SOCP} and \ac{SDR}. An optimal linear precoder was employed in \cite{Asymptotic_Max-Min_SINR_RIS} to maximize the minimum \ac{SINR} under a power constraint. In \cite{max_min_RIS_NOMA}, a max-min fairness algorithm was developed for optimizing active-passive beamforming and power allocation in \ac{RIS}-assisted non-orthogonal multiple access systems, using block coordinate descent and limited-memory Broyden-Fletcher-Goldfarb-Shanno algorithms. The work in \cite{max_min_RIS_cell_free2} considered max-min rate optimization for \ac{RIS}-assisted cell-free systems by optimizing transmit power and phase shifts with \ac{GP} and \ac{SDR}. In \cite{Max_min_RIS_green}, max-min \ac{SINR} optimization was investigated under users’ maximum allowable electromagnetic exposure. In addition, \cite{max_min_RIX_convex_hull_relaxation} proposed a convex-hull relaxation for the discrete phase shift constraint and solved the reformulated problem using an alternating projection/proximal gradient descent and ascent algorithm. Finally, \cite{max_min_RIS_GDA} applied a  \ac{GDA}–based approach along with \ac{SDR} to solve the max-min \ac{SNR} problem in a multi-user \ac{RIS}-assisted system.

To reduce computational overhead, several studies employed statistical \ac{CSI}–based optimization. In \cite{SIM_optimization_statistical_CSI}, transmit power and \ac{RIS} phase shifts were optimized based on statistical \ac{CSI} to improve rate performance in \ac{SIM}-assisted \ac{MIMO} systems, using the use-and-then-forget bound for downlink spectral efficiency. The work in \cite{max_min_RIS_cell_Free} derived an approximate closed-form expression for the achievable uplink net rate of a \ac{RIS}-assisted cell-free system using statistical \ac{CSI}, then solved the max-min \ac{SINR} problem by jointly optimizing receiver filter coefficients, power allocations, and \ac{RIS} phase shifts through closed-form solutions, \ac{GP}, and alternating maximization. In \cite{max_min_star_RIS}, the performance of a simultaneously transmitting and reflecting \ac{RIS} in a massive \ac{MIMO} downlink system with hardware impairments was examined using an optimal linear precoder and a passive beamforming matrix to maximize the minimum \ac{SINR}, with amplitude and phase shifts optimized via a projected gradient ascent algorithm based on statistical \ac{CSI}. Similarly, \cite{S_CSI_RIS_optimization} investigated transmit power allocation and phase shift optimization based on statistical \ac{CSI} while ensuring user fairness by maximizing the sum of eigenvalues of the composite channel’s correlation matrix through an iterative algorithm and \ac{GP}. An upper bound on the ergodic spectral efficiency for \ac{RIS}-assisted systems was derived in \cite{RIS_statisticalCSI}, with \ac{RIS} phase shifts optimized using \ac{SDR} and Gaussian randomization. In \cite{RIS_statisticalCSI2}, an approximate closed-form expression for the uplink ergodic data rate was derived and a genetic algorithm was used for phase shift optimization. Finally, \cite{RIS_statisticalCSI_DRL} obtained an approximate closed-form expression of the ergodic sum rate and addressed the joint active-passive beamforming problem using deep reinforcement learning.

\subsection{MOTIVATION AND CONTRIBUTIONS}

As discussed, existing studies on \ac{SIM}-assisted wireless networks have primarily focused on maximizing the achievable sum rate or spectral efficiency. To the best of our knowledge, the max-min fairness problem in these systems remains unexplored. Ensuring fairness is crucial to guarantee that all users receive a reasonable quality of service, especially in multi-user scenarios. Additionally, optimizing resource allocation based on statistical \ac{CSI} reduces computational overhead by eliminating the need for frequent re-optimization. However, this approach has yet to be investigated in the context of max-min fairness for \ac{SIM}-assisted systems.
Motivated by this gap, we develop a max-min fairness algorithm for \ac{SIM}-assisted multi-user \ac{MISO} systems that jointly optimizes power allocation and wave-based beamforming using both instantaneous and statistical \ac{CSI}. Our main contributions are as follows:

% In the mathematical optimization-related literature for \ac{SIM}-assisted wireless networks, investigations on the maximization of achievable sum rate or spectral efficiency are only available \cite{an2023stackedintelligentmetasurfacesmultiuser,SIM_optimization_holographic_MIMO, SIM_optimization_statistical_CSI,SIM_fully_analog_beamforming,SIM_optimization_DRL}. However,  finding optimal resource allocation by maximizing the minimum rate or \ac{SINR} of the \ac{SIM}-assisted wireless network is crucial to guarantee fairness. Moreover, optimizing based on statistical \ac{CSI} helps to reduce the computational overhead by removing the need for frequent optimization. Even though it is an important task, the problem of max-min fairness optimization, either based on instantaneous \ac{CSI} or statistical \ac{CSI}, has not yet been investigated. This literature gap motivates us to find a max-min fairness algorithm to find optimal power allocation and optimal wave-based beamforming in \ac{SIM}-assisted multi-user \ac{MISO} systems based on both instantaneous \ac{CSI} and statistical \ac{CSI}. The main contributions of
% this work are further elaborated in the following. 
\begin{itemize}
    \item This paper proposes two novel max-min fairness algorithms for \ac{SIM}-assisted multi-user \ac{MISO} downlink systems. One algorithm is designed for instantaneous \ac{CSI}, while the other is based on statistical \ac{CSI}. Both approaches iteratively optimize power allocation and wave-based beamforming to improve fairness among users. Specifically, we assume the perfect \ac{CSI} knowledge in every coherence interval during optimization with instantaneous \ac{CSI} and the knowledge of channel characteristics for a selected period of coherence intervals during optimization with statistical \ac{CSI}.
    \item For the instantaneous \ac{CSI} case, we formulate an optimization problem to maximize the minimum \ac{SINR}, subject to the total transmit power budget and the practical constraint of discrete phase shifts. To solve this problem, we propose an alternating optimization algorithm that iteratively updates power allocation and wave-based beamforming. Specifically, \ac{GP} is used to refine power allocation for a given wave-based beamforming matrix, while a \ac{GDA}-based method adjusts the beamforming for a fixed power allocation.
    \item For the statistical \ac{CSI} case, we formulate an optimization problem based on a bound on the average minimum achievable rate, as the exact average rate is analytically intractable. To address this, we develop an alternating optimization algorithm that combines geometric programming for power allocation with a gradient-descent-based method for wave-based beamforming.
    \item Finally, we provide simulation results to evaluate the performance of the proposed max-min fairness optimization algorithms. The results show that our approach achieves an order of magnitude higher minimum rate and higher fairness compared to equal power allocation and random phase shift assignment with instantaneous \ac{CSI}. Moreover, while both power allocation and wave-based beamforming improve performance, the latter has a more significant impact. Furthermore, the results confirm that the proposed upper bound on the average minimum achievable rate remains tight in the low-SNR regime with statistical \ac{CSI}.
\end{itemize}

The rest of the paper is organized as follows. The system model for the SIM-assisted system is described in Section \ref{SM}. In Section \ref{max-min-sinr}, we present the proposed max-min fairness optimization algorithm based on instantaneous CSI, while in Section \ref{max-min-erate}, we present the proposed max-min fairness optimization algorithm based on statistical CSI. The numerical results are presented in Section \ref{results}, and Section \ref{conclusion} concludes our paper.
%%%%%%%%%%%%%%%%%%%%%%%%%%%%%%%%%%%%%%%%%%%%%%%%%%%%%
\subsubsection*{Notations}
Lowercase letters represent scalars, boldface lowercase letters denote column vectors, and boldface uppercase letters represent matrices.  Moreover, $|z|$, $\Re(z)$, and $\Im(z)$ denote the absolute value, real part, and imaginary part of a complex number $z$, respectively. $\mathsf{j}$ is the imaginary unit which satisfies
$\mathsf{j}^2 = -1$. The notations for the transpose, the Hermitian transpose and the Euclidean norm of $\mathbf{A}$ are given by $\mathbf{A}\tran, \mathbf{A}\herm $, and $||\mathbf{A}||_2$, respectively. The operator $\diag(\cdot)$ transforms a vector into a diagonal matrix. $\mathbf{I}_N$ is the $N\times N$ identity matrix. The square root of a positive-definite/ positive-semi-definite matrix is denoted by $(\cdot)^{1/2}$. The mathematical expectation is denoted by $\mathbb{E}\left\{\cdot\right\}$. Additionally, $\mathbf{1}$ and $\mathbf{0}$ represent the all-one vector and the all-zero vector, respectively.  Moreover, $\mathrm{sinc}(x) = \frac{\sin{(\pi x)}}{\pi x}$ is the  sinc function and $\frac{\partial f}{\partial {x}}$ is the partial derivative of a function $f$ with respect to $x$. Finally, $\lfloor \cdot \rfloor$ represents the floor operator and $\log_a(\cdot)$ is the logarithmic function with base $a$.

%%%%%%%%%%%%%%%%%%%%%%%%%%%%%%%%%%%%%%%%%
\section{SYSTEM MODEL}\label{SM}
\begin{figure}[t]
    \centering
        \includegraphics[width=1\columnwidth]{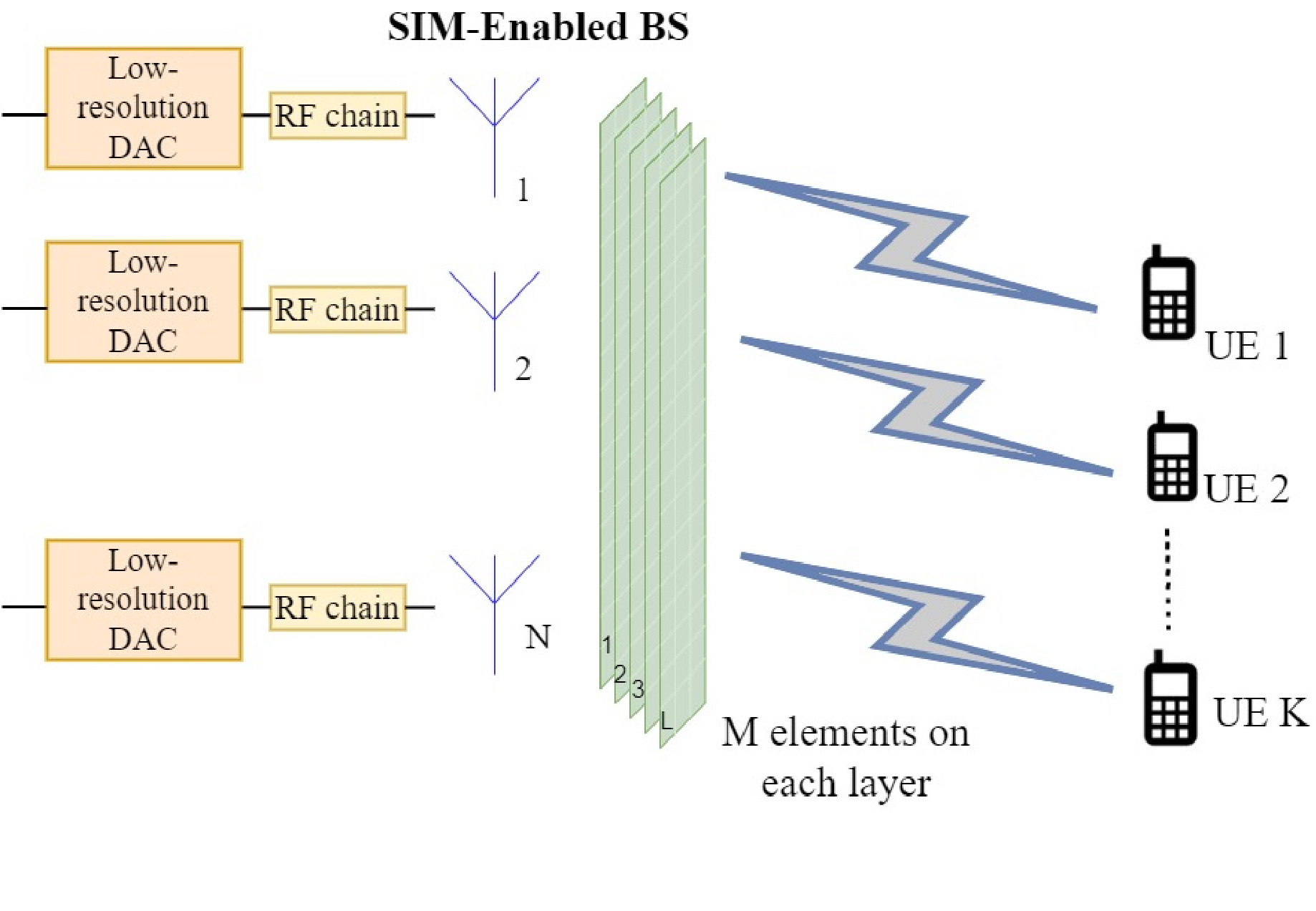 }
  \caption{SIM-assisted communication system.}
  \label{fig:Illustration of the system model}
 \vspace{-2mm}
\end{figure}

%SIM is an emerging technology that uses multiple \ac{RIS} layers to perform wave-based beamforming. 
We consider in this work a downlink \ac{SIM}-assisted multi-user \ac{MISO} system, as shown in Fig.~\ref{fig:Illustration of the system model}, where one \ac{BS} employing $N$ transmit antennas is assisted by a \ac{SIM} and communicates with $K$ single-antenna \acp{UE}.  The \ac{SIM} is made with $L$ metasurface layers and positioned near the \ac{BS} antennas. Each SIM layer comprises $M$ scattering elements. Let $\mathcal{L}=\{1, \cdots, L\}$, $\mathcal{M} = \{1, \cdots, M\}$ and $\mathcal{K} = \{1, \cdots, K\}$ denote the sets of metasurface layers, scattering elements on each layer, and \ac{UE}s, respectively.

The phase shift of the $m$-th scattering element on the ${\ell}$-th metasurface layer is denoted as $\phi_m^{(\ell)}=\alpha_m^{(\ell)} e^{\mathsf{j}\theta_m^{(\ell)}}$, where $\alpha_m^{(\ell)}$ and $\theta_m^{(\ell)}$ are the amplitude and phase shift, respectively, such that $\alpha_m^{(\ell)}= 1$ and $\theta_m^{(\ell)} \in[0,2 \pi), \forall m \in \mathcal{M}, \forall \ell \in \mathcal{L} $. Moreover, we assume discrete phase shifts, where the interval $[0,2 \pi)$ is uniformly quantized into $2^b$ levels, with $b$ denoting the number of quantization bits. The set of discrete phase shifts is defined as $\mathcal{B} = \{0,\Delta_\theta,2\Delta_\theta, \cdots,(2^b-1)\Delta_\theta \}$, where $\Delta_\theta = 2\pi/2^b$.
Thus, the phase-shift $\theta_m^{(\ell)}$ is quantized to the nearest values in $\mathcal{B}$ as follows:
 \begin{align}
     \theta_m^{(\ell)} &\leftarrow  \text{arg} \min_{\theta \in \mathcal{B}} |\theta-\theta_m^{(\ell)}|,\notag \\
      &= \Delta_\theta \left\lfloor \frac{\theta_m^{(\ell)}}{\Delta_\theta} + \frac{1}{2} \right\rfloor, \forall m \in \mathcal{M}, \forall \ell \in \mathcal{L} \label{quantization1}.
 \end{align}
Moreover, the phase shifts matrix of the ${\ell}$-th metasurface layer is defined as 
\begin{equation}
    \bm{\Theta}_{\ell} = \diag\left(e^{\mathsf{j}\theta_1^{(\ell)}}, e^{\mathsf{j}\theta_2^{(\ell)}},\cdots,e^{\mathsf{j}\theta_M^{(\ell)}}\right) \in \mathbb{C}^{M \times M}.
\end{equation}

According to Rayleigh-Sommerfeld's diffraction theory\cite{SIM_intro2,doi:10.1126/science.aat8084}, we can obtain the propagation channel matrix between the $({\ell}-1)$-th layer and ${\ell}$-th layer which is denoted by $\mathbf{W}^{({\ell})} \in \mathbb{C}^{M \times M}, {\ell} = 2,\cdots ,L$. The $(i,j)$-th entry of  $\mathbf{W}^{({\ell})}$ can be written as 
\begin{equation}
    w^{({\ell})}_{i,j} = \frac{\rm{d}_x\rm{d}_y\rm{d}_{\rm{Layer}}}{{\rm{d}_{i,j}^{({\ell})}}^2}\left( \frac{1}{2\pi \rm{d}_{i,j}^{({\ell})}}-\frac{1}{\lambda}\mathsf{j}\right)e^{\mathsf{j}2\pi d^{({\ell})}_{i,j}/\lambda}, \label{lthchannel}
\end{equation}
where $\lambda $ is the wavelength, $\rm{d}_x$ and $\rm{d}_y$ are the length and width of each \ac{RIS} element,  and $\rm{d}^{({\ell})}_{i,j}$ is the transmission distance between $j$-th scattering element of $({\ell}-1)$-th layer to the $i$-th scattering element of the ${\ell}$-th layer. We assume the spacing between each layer is the same and it is denoted by  $\rm{d}_{\rm{Layer}}$.

The propagation channel matrix from \ac{BS} to the first metasurface layer is denoted by $\mathbf{W}^{(1)} =[\mathbf{w}_1,\cdots, \mathbf{w}_N ] \in \mathbb{C}^{M \times N}$ and can be obtained by replacing $\rm{d}^{({\ell})}_{i,j}$ with $\rm{d}^{(1)}_{m,n}$ and $\rm{d}_{\rm{Layer}}$ with $\rm{d}_t$, where $\rm{d}_t$ is the distance between the \ac{BS} and the first metasurface layer \cite{SIM_intro2}. The wave-based beamforming matrix of the \ac{SIM}-assisted system can be written as 
\begin{equation}
    \mathbf{G}_{\bm{\vartheta}} = \left(  \prod_{k=1}^{L-1} \bm{\Theta}_{L+1-k}\mathbf{W}^{(L+1-k)} \right) \bm{\Theta}_{1} \in \mathbb{C}^{M \times M}.
\end{equation}
The channel matrix from the last metasurface layer to the \ac{UE}s is denoted by $\mathbf{H}=[\mathbf{h}_1,\cdots, \mathbf{h}_K] \in \mathbb{C}^{M \times K}$, where $\mathbf{h}_K\herm \in \mathbb{C}^{1\times M}, \forall \mathcal{K}$  is the \ac{SIM}-UE channel vector which can be written as 
\begin{equation}\mathbf{h}_k =\sqrt{\beta_{k}}{\mathbf {R}}_{\text{RIS}}^{1/2}{\mathbf {q}}_{h_k},\end{equation}
where ${\mathbf {q}}_{h_k}  \sim {\mathcal{CN}}( {0,\mathbf{1}_M}) $ follows the Rayleigh fading distribution %({\color{red} It is not clear what you mean by Rayleigh here.})
and $\sqrt{\beta_{k}}$ is the distance-dependent path loss corresponding to the \ac{SIM}-UE channels. Furthermore, we assume far-field propagation in an isotropic scattering environment and  ${\mathbf {R}}_{\text{RIS}} \in {\mathbb{C}^{M \times M}}$ denotes the deterministic Hermitian-symmetric positive semi-definite correlation matrix at the \ac{RIS} in which the $(i,j)$-th term is given by
\begin{equation}
    {\mathbf {R}}_{\text{RIS}_{i,j} }= \mathrm{sinc} \left(\frac{2\sqrt{{\rm{d}^{(H)}_{i,j}}^2+{\rm{d}^{(V)}_{i,j}}^2}}{\lambda} \right),
\end{equation}
where $\rm{d}^{(H)}_{i,j}$ and $\rm{d}^{(V)}_{i,j}$ represent the horizontal and vertical distances, respectively, between the $m$-th and  $j$-th  meta-atoms located on the same metasurface layer \cite{SIM_intro2,SIM_intro}.

In downlink transmission, we assume that each data stream related to each \ac{UE} transmits from a corresponding \ac{BS} antenna, which means the \ac{BS} needs to select a suitable number of \ac{BS} antennas in advance for the $\mathcal{K}$ \ac{UE} set. For simplicity, we assume $N=K$. Specifically, the $k$-th \ac{UE} treats signals intend for the other \ac{UE}s as interference. However, we use wave-based beamforming with the aid of \ac{SIM}. Thus, we can adjust $\mathbf{G}$ by optimizing phase shifts to mitigate interference. Let $s_k\in \mathbb{C}$ be the information symbol intended for the $k$-th \ac{UE} with $\mathbb{E}\left\{||s_k||^2\right\}=1$, i.e., unit-energy constellation, and $p_k > 0$ be the power allocation for the $k$-th \ac{UE}. Therefore, total power constraint can be expressed as follows
\begin{equation}
    \sum_{k=1}^K p_k \leq P_T,
\end{equation} where $P_T$ is the power budget at the \ac{BS}. The composite received signal at the $k$-th \ac{UE} after superimposing all the signals through \ac{SIM}  assumes
\begin{equation}
    r_k  = \sqrt{p_k}\mathbf{h}_k\herm \mathbf{G}_{\bm{\vartheta}}\mathbf{w}_{k}s_{k} +\sum_{\substack{j=1 \\ j \neq k}}^K\sqrt{p_j}\mathbf{h}_k\herm \mathbf{G}_{\bm{\vartheta}}\mathbf{w}_{j}s_{j}+ n_k, \forall k \in \mathcal{K},
\end{equation} where $n_k  \sim {\mathcal{CN}}( {0,{\sigma^2}})$ denotes the complex white Gaussian noise.  

%%%%%%%%%%%%%%%%%%%%%%%%%%%%%%%%%%%%%%%%%%%%%%%%%%%%%%%%%%%%%%%%%%%%%%%%%%%%%%%%%%%%%%%%%%%%%%%%%%%%%%%%%%%%%%%%%%%%
\section{JOINT POWER ALLOCATION AND WAVE-BASED BEAMFORMING WITH THE INSTANTANEOUS CSI}
\label{max-min-sinr}
\subsection{PROBLEM FORMULATION}
In this section, our objective is to maximize the minimum per-UE achievable rate by jointly optimizing the transmit power allocation and the wave-based beamforming at the \ac{SIM} under instantaneous \ac{CSI}. The \ac{SINR} at $k$th \ac{UE} can be expressed as
\begin{equation}
    \gamma_k = \frac{|{\mathbf{h}_k\herm} \mathbf{G}_{\bm{\vartheta}}\mathbf{w}_{k}|^2p_k}{\sum_{\substack{j=1 \\ j \neq k}}^K|\mathbf{h}_k\herm \mathbf{G}_{\bm{\vartheta}}\mathbf{w}_j|^2p_j + \sigma^2}, \forall  k \in \mathcal{K}.
\end{equation}
Therefore, the achievable rate of the $k$-th \ac{UE} is
\begin{equation}
    R_k = \log_2(1 +    \gamma_k ), \forall k \in \mathcal{K}.
\end{equation}
Since $\log_2(1 + \gamma_k )$ is a monotonically increasing function of $\gamma_k\geq 0$, the corresponding optimization problem can be formulated as 
\begin{subequations}
\begin{align}
    (P 1):\ \max _{\mathbf{p}, \boldsymbol{\vartheta}} {\min_{k \in \mathcal{K}}}  & \quad      \frac{|{\mathbf{h}_k\herm} \mathbf{G}_{\bm{\vartheta}}\mathbf{w}_{k}|^2p_k}{\sum_{\substack{j=1 \\ j \neq k}}^K|\mathbf{h}_k\herm \mathbf{G}_{\bm{\vartheta}}\mathbf{w}_j|^2p_j + \sigma^2} ,   \\
    \mbox{subject to} 
      &\quad \sum_{k=1}^{K} p_{k} \leq P_{T}, %\label{p1_b}
    \\ & \quad p_{k} > 0, \forall k \in \mathcal{K}, %\label{p1_c}
    \\ &\quad  \theta_m^{(\ell)} \in \mathcal{B},  \forall m \in \mathcal{M}, \forall \ell \in \mathcal{L},
   % \label{p1_d}
\end{align}
\end{subequations}
where $\mathbf{p}= [p_1, p_2, \cdots, p_K]\tran \in \mathbb{R}^{K \times 1}$, $\bm{\theta}_{\ell} = [\theta_1^{(\ell)}, \theta_2^{(\ell)},\cdots,\theta_M^{(\ell)}]\tran\in \mathbb{R}^{M \times 1}$ and $\bm{\vartheta} = [ \bm{\theta}_1\tran, \bm{\theta}_2\tran, \cdots, \bm{\theta}_L\tran]\tran \in \mathbb{R}^{ML \times 1}$.

The optimization problem (\emph{P}1) is non-convex due to the coupling of $\mathbf{p}$ and $\bm{\vartheta}$ in its objective function and restricted discrete values for $\theta_m^{(\ell)}$ in the constraint \ref{p1_d}. Thus, standard methods cannot be used to find a globally optimal solution. To address this, we propose an algorithm which decomposes the problem (\emph{P}1) into two sub-problems. Additionally, we propose to solve the above-mentioned two sub-problems alternatively to obtain a sub-optimal solution described in the section \ref{AO}.

\subsection{THE PROPOSED ALTERNATING OPTIMIZATION ALGORITHM}
\label{AO}  
In this section, we describe the proposed alternating optimization algorithm for solving the max-min \ac{SINR} optimization problem. Specifically, we solve the power allocation problem in the first sub-problem and the wave-based beamforming problem in the second sub-problem, as listed below. 
\begin{itemize}
    \item  The power allocation optimization problem for fixed \ac{SIM} phase shifts can be formulated as
\begin{subequations}
\begin{align}
    (P 2):\ \max _{\mathbf{p}} {\min_{k \in \mathcal{K}}}  & \quad      \frac{|{\mathbf{h}_k\herm} \mathbf{G}_{\bm{\vartheta}}\mathbf{w}_{k}|^2p_k}{\sum_{\substack{j=1 \\ j \neq k}}^K|\mathbf{h}_k\herm \mathbf{G}_{\bm{\vartheta}}\mathbf{w}_j|^2p_j + \sigma^2} ,   \\
    \mbox{subject to} 
     &\quad  \sum_{k=1}^{K} p_{k} \leq P_{T},\label{p1_b}
    \\ & \quad p_{k} > 0, \forall k \in \mathcal{K}. \label{p1_c}
     %\mbox{constraints~}  \eqref{p1_b}, \eqref{p1_c} .
\end{align}
\end{subequations}
 \item The optimal wave-based beamforming problem for a given power allocation can be formulated as 
\begin{subequations}
\begin{align} 
        (P 3):\ \max _{ \boldsymbol{\vartheta}} {\min_{k \in \mathcal{K}}}  & \quad      \frac{|{\mathbf{h}_k\herm} \mathbf{G}_{\bm{\vartheta}}\mathbf{w}_{k}|^2p_k}{\sum_{\substack{j=1 \\ j \neq k}}^K|\mathbf{h}_k\herm \mathbf{G}_{\bm{\vartheta}}\mathbf{w}_j|^2p_j + \sigma^2} ,   
    \\ \mbox{subject to} &\quad  \theta_m^{(\ell)} \in \mathcal{B},  \forall m \in \mathcal{M}, \forall \ell \in \mathcal{L}.
    \label{p1_d}%\mbox{constraints~}  \eqref{p1_d} .
\end{align}    
\end{subequations}
\end{itemize}
Thus, we propose a two-stage algorithm to solve the optimization problem (\emph{P}1) as presented in the algorithm \ref{alg:1}. 
 
\begin{algorithm}[t]
\caption{Max-Min \ac{SINR} Optimization.}\label{alg:1}
\begin{algorithmic}[1]
\STATE \textbf{Initialization:}  Set $i =1$. Initialize the \ac{SIM} phase shifts $\bm{\theta}_\ell  = \mathbf{0}, \forall \ell \in \mathcal{L}$. 
\STATE Solve the optimization problem (\emph{P}2) and find the optimal power allocation for the $i$-th iteration for the given $\bm{\theta}_\ell \forall \ell \in \mathcal{L}$. 
\STATE Solve the optimization problem (\emph{P}3)  and find the optimal \ac{SIM} phase shifts 
for the $i$-th iteration for the given $\mathbf{P}$.
\STATE Repeat steps  2 and 3 until convergence.
\STATE $i=i+1$.
\end{algorithmic}
\end{algorithm}

%%%%%%%%%%%%%%%%%%%%%%%%%%%%%%%%%%%%%%%%%%%%%%%%%%%%%%%%%%%%%%%%%%%%%%%%%%%%%%%%%%%%%%%%%%%%%%%%%%%%%%%%%%%%%%%%%%
\subsubsection{Optimal Power Allocation With Given \boldmath $\vartheta$}

Specifically, we can reformulate the optimization problem (\emph{P}2) in epigraph form by introducing an auxiliary variable $t$ as follows \cite{boyd2004convex},
\begin{subequations}
\begin{align} 
        (P 2.1):\ \max _{t \geq 0, \mathbf{p}}  & \quad t
    \\ \mbox{subject to} &\quad t \leq   \frac{|{\mathbf{h}_k\herm} \mathbf{G}_{\bm{\vartheta}}\mathbf{w}_{k}|^2p_k}{\sum_{\substack{j=1 \\ j \neq k}}^K|\mathbf{h}_k\herm \mathbf{G}_{\bm{\vartheta}}\mathbf{w}_j|^2p_j + \sigma^2},\forall k\in\mathcal{K}, 
    \\ &\quad \mbox{constraints~}  \eqref{p1_b}, \eqref{p1_c}.
\end{align}    
\end{subequations}
The problem (\emph{P}2.1) can be rewritten as 
\begin{subequations}
\begin{align} 
        (P 2.2):\ \max _{t \geq 0, \mathbf{p}}  & \quad t
    \\ \mbox{subject to} &\quad {\sum_{\substack{j=1 \\ j \neq k}}^Ks_{k,j}p_j + \sigma^2_k} \leq \frac{s_{k,k}}{t} ,\forall k\in\mathcal{K}, 
    \\ &\quad \mbox{constraints~}  \eqref{p1_b}, \eqref{p1_c}.
\end{align}    
\end{subequations}
where $s_{k,k}=|\mathbf{h}_k\herm \mathbf{G}_{\bm{\vartheta}}\mathbf{w}_j|^2$. Thus, the objective function and the constraints are monomial and polynomial functions. Hence,
 the optimization problem (\emph{P}2.2) follows the standard form of the \ac{GP} problem \cite{boyd2004convex}, and it can be solved using standard optimization tools.

%%%%%%%%%%%%%%%%%%%%%%%%%%%%%%%%%%%%%%%%%%%%%%%%%%%%%%%%%%%%%%%%%%%%%%%%%%%%%%%%%%%%%%%%%%%%%%%%%%%%%%%%%%%%%%%%%%%%%%
\subsubsection{Optimal Wave-Based Beamforming With Given \boldmath$p$}

We propose a \ac{GDA}-based algorithm to solve the wave-based beamforming optimization problem described below. Here, we relaxed the discrete phase shifts into the continuous domain.  Thereby, we transform the optimization problem (\emph{P}3), by introducing an auxiliary vector variable $\bm{\lambda} = [\lambda_1, \cdots , \lambda_K]\tran \in \mathbb{R}^{K \times 1}$ and $\Lambda =\{\bm{\lambda}| \lambda_k \geq 0, \forall k, \mathbf{1}\tran\bm{\lambda} =1 \}$ as follows:
\begin{subequations}
\begin{align} 
        (P 3.1):\ \max _{ \boldsymbol{\vartheta}} {\min_{\bm{\lambda}}}  & \quad      \sum_{k=1}^K{\frac{\lambda_k|{\mathbf{h}_k\herm} \mathbf{G}_{\bm{\vartheta}}\mathbf{w}_{k}|^2p_k}{\sum_{\substack{j=1 \\ j \neq k}}^K|\mathbf{h}_k\herm \mathbf{G}_{\bm{\vartheta}}\mathbf{w}_j|^2p_j + \sigma^2}} ,
        \\  \mbox{subject to} &\quad \bm{\lambda} \geq 0, \mathbf{1}\tran\bm{\lambda} =1 ,
        \\ &\quad  \mbox{constraints~}   \eqref{p1_d}.
\end{align}    
\end{subequations}
The equivalence of Problem (\emph{P}3) and Problem (\emph{P}3.1) arises from the fact that, for a fixed $\bm{\vartheta}$, we assign $\lambda_k=1$ for the minimum and $\lambda_i=0, i \neq k$ for the rest \cite{max_min_RIS_GDA}.

Let us define $f(\bm{\lambda},\bm{\vartheta})= \sum_{k=1}^K{\frac{\lambda_k|{\mathbf{h}_k\herm} \mathbf{G}_{\bm{\vartheta}}\mathbf{w}_{k}|^2p_k}{\sum_{\substack{j=1 \\ j \neq k}}^K|\mathbf{h}_k\herm \mathbf{G}_{\bm{\vartheta}}\mathbf{w}_j|^2p_j + \sigma^2}}  .$  Therefore, for fixed $\bm{\vartheta}$, the auxiliary variable $\bm{\lambda}$ can be updated by applying a projection gradient descent step as follows:
\begin{subequations}  \label{lambda_update}
\begin{align}
    \bar{\bm{\lambda}}^{(i+1)} &= \bm{\lambda}^{(i)} -  \epsilon\cdot{\nabla_{\bm{\lambda}} f(\bm{\lambda}^{(i)},\bm{\vartheta}^{(i)})}, \\
    \bm{\lambda}^{(i+1)} &= \arg \min_{\mathbf{y} \in S_{\bm{\lambda}}}|| \mathbf{y}- \bar{\bm{\lambda}}^{(i+1)} ||^2_2, \label{proj_lambda}
    \end{align}
\end{subequations}
where $\epsilon$ is the step size and $S_{\bm{\lambda}} =\{\mathbf{y}| \mathbf{y} \in \mathbb{R}^{K \times 1}, \mathbf{y} \geq 0, \mathbf{1}_K\tran\mathbf{y} =1 \}$. Moreover ${\nabla_{\bm{\lambda}} f} = \left[ \frac{\partial f}{\partial {\lambda_1}} , \cdots,\frac{\partial f}{\partial {\lambda_K}}\right]\tran$  is the gradient of $f$ with respect to $\bm{\lambda}$ where $ \frac{\partial f}{\partial {\lambda_k}}$ can obtained as follows:
\begin{equation}
    \frac{\partial f}{\partial {\lambda_k}} = \sum_{k=1}^K { \frac{\partial \lambda_k\gamma_k}{\partial {\lambda_k }} }=\gamma_k .
\end{equation}
The projection operation in \eqref{proj_lambda} can be effectively solved using the \ac{KKT} conditions since the optimization problem \eqref{proj_lambda} exhibits convexity. Therefore, its Lagrangian function can be written as follows.
\begin{equation}
    L(\mathbf{y},\bm{\xi},\eta)= || \mathbf{y}- \bar{\bm{\lambda}}^{(i+1)} ||^2_2 + \bm{\xi}\tran(-\mathbf{y}) +\eta(\mathbf{1}_K\tran\mathbf{y} -1), \label{dual_problem}
\end{equation}
where $\bm{\xi} \in \mathbb{R}^{K \times 1} $ and $\eta \in \mathbb{R}$ are the Lagrange multipliers. We can obtain the \ac{KKT} conditions as listed below.
\begin{itemize}
    \item $\nabla_{\mathbf{y}}( L(\mathbf{y},\bm{\xi}^*,\eta^*))$ vanishes at $ \mathbf{y^*}$: 
        \begin{equation}
            \mathbf{y^*} = \bar{\bm{\lambda}}^{(i+1)} + \frac{1}{2}(\bm{\xi}^* - \eta^*\mathbf{1}_K). 
        \end{equation}
    \item {Prime feasibility:} 
        \begin{subequations}
        \begin{align}
             \bar{\bm{\lambda}}^{(i+1)}  &\geq \frac{1}{2}(\eta^*\mathbf{1}_K - \bm{\xi}^*  ), \label{prime1} \\
             K\eta^* &= \mathbf{1}_K\tran(\bm{\xi}^* + 2 \bar{\bm{\lambda}}^{(i+1)} ) -2.\label{prime2}
        \end{align} 
        \end{subequations}
    \item Dual feasibility:
        \begin{equation}
            \bm{\xi}^* \geq 0. \label{dual}
        \end{equation}
    \item Complementary slackness: Using \eqref{prime1} and \eqref{dual},
        \begin{equation}
              \bm{\xi}^* = \max(\eta^*\mathbf{1}_K-2 \bar{\bm{\lambda}}^{(i+1)} , \mathbf{0}). \label{CS}
        \end{equation}
\end{itemize}
Here, $ \bm{\xi}^*$,$\eta^* $, and $\mathbf{y^*}$ are the optimal solutions for the problem \eqref{dual_problem}. With the help of \eqref{CS}, we can define $ \bm{\xi}^*+2 \bar{\bm{\lambda}}^{(i+1)} = \max(\eta^*\mathbf{1}_K,2 \bar{\bm{\lambda}}^{(i+1)}) $. According to \eqref{prime2}, we can further define 
\begin{equation}
    g(\eta) =  \mathbf{1}_K\tran.\max(\eta^*\mathbf{1}_K,2 \bar{\bm{\lambda}}^{(i+1)}) -2.
\end{equation}
$ g(\eta)$ is monotonic in $\eta$ and $\frac{2}{K}\mathbf{1}_K\tran\bar{\bm{\lambda}}^{(i+1)} - \frac{2}{K}  \leq \eta^* \leq  2\max_k\bar{\bm{\lambda}}^{(i+1)}_k   $. Therefore,  we can find a solution for $\eta^* $ using bisection search until predefined accuracy $\varepsilon$ is achieved. Thus, for a given $\eta^* $, we can find optimal  $ \bm{\xi}^*$ and $\mathbf{y^*}$ and then we can update $\bm{\lambda}^{(i+1)}$.

Moreover, for fixed $\lambda$, all the \ac{SIM} phase shifts $\bm{\vartheta}$ can be updated using gradient ascent step as follows:
\begin{subequations}\label{psi_update}
\begin{align}
    \bar{{\bm{\vartheta}}}^{(i+1)} &=  {\bm{\vartheta}}^{(i)} +  \mu\cdot{\nabla_{\bm{\vartheta}}  f(\bm{\lambda}^{(i+1)},\bm{\vartheta}^{(i)})},\\
    {\bm{\vartheta}}^{(i+1)} &= \arg\min_{\mathbf{b} \in S_{\bm{\vartheta}}}|| \mathbf{x} - \bar{\bm{\vartheta}}^{(i+1)} ||^2_2,
 \end{align}
\end{subequations}
where $S_{\bm{\vartheta}}= \{\mathbf{x}|\mathbf{x} \in \mathbb{R}^{ML \times 1}, x_i \in[0,2 \pi) \}$. Here, $\mu$ is the step size which is updated using a backtracking line search \cite{boyd2004convex}. 
Specifically, ${\nabla_{\bm{\vartheta}} f} = \left[ \frac{\partial f}{\partial {\theta_1^{(1)}}} ,\cdots,\frac{\partial f}{\partial {\theta_M^{(1)}}},\cdots,\frac{\partial f}{\partial {\theta_1^{(L)}}}, \cdots,  \frac{\partial f}{\partial {\theta_M^{(L)}}}\right]\tran$ is the gradient of $f$ with respect to $\bm{\vartheta}$, where $\frac{\partial f}{\partial {\theta_m^{(\ell)}}}$ is obtained in the following lemma. %({\color{red}It seems you have relaxed the discrete nature of $\theta_k$s to obtain the continuous analogy so that the partial derivatives are meaningful. I am not sure whether you have alreday spelled it out. }) 
% \begin{align}
%     \frac{\partial f}{\partial {\theta_m^{(\ell)}}} = \sum_{k=1}^K{\lambda_k\omega_k (p_k\delta_{m,k,k}^{(\ell)} - \gamma_k\sum_{\substack{j=1 \\ j \neq k}}^K p_j\delta_{m,k,j}^{(\ell)})},
% \end{align}
% where $\delta_{m,k,i}^{(\ell)}$ and $ \omega_k$ is defined as
% \begin{align}
%  \delta_{m,k,j}^{(\ell)} &=2\Im\left[ e^{-\mathsf{j}\theta_m^{(\ell)}}\mathbf{w}_j\herm\mathbf{a}^\ell_m(\mathbf{b}^\ell_m)\herm\mathbf{h}_k\mathbf{h}_k\herm \mathbf{G}_{\bm{\vartheta}}\mathbf{w}_j\right] ,\\
%  \omega_k &= \frac{1}{\sum_{\substack{j=1 \\ j \neq k}}^K|\mathbf{h}_k\herm \mathbf{G}_{\bm{\vartheta}}\mathbf{w}_j|^2p_j + \sigma^2},
% \end{align}
% respectively.

\paragraph*{Lemma I}\label{lemma:gradient}
The gradient of the objective function $f$ with respect to the phase shift $\theta_m^{(\ell)}$ at the $m$-th element of the $\ell$-th metasurface layer is given by:
\begin{align}
\frac{\partial f}{\partial \theta_m^{(\ell)}} = \sum_{k=1}^K \lambda_k \omega_k \left( p_k \delta_{m,k,k}^{(\ell)} - \gamma_k \sum_{\substack{j=1 \\ j \neq k}}^K p_j \delta_{m,k,j}^{(\ell)} \right),
\end{align}
where $\delta_{m,k,j}^{(\ell)}$ and $\omega_k$ are defined as
\begin{align}
 \delta_{m,k,j}^{(\ell)} &=2\Im\left[ e^{-\mathsf{j}\theta_m^{(\ell)}}\mathbf{w}_j\herm\mathbf{a}^\ell_m(\mathbf{b}^\ell_m)\herm\mathbf{h}_k\mathbf{h}_k\herm \mathbf{G}_{\bm{\vartheta}}\mathbf{w}_j\right] ,\\
 \omega_k &= \frac{1}{\sum_{\substack{j=1 \\ j \neq k}}^K|\mathbf{h}_k\herm \mathbf{G}_{\bm{\vartheta}}\mathbf{w}_j|^2p_j + \sigma^2},
\end{align}
with $(\mathbf{a}_m^\ell)\herm$ and $\mathbf{b}_m^\ell$ denoting the $m$-th row of $\mathbf{A}_\ell \in \mathbb{C}^{M \times M}$ and the $m$-th column of $\mathbf{B}_\ell \in \mathbb{C}^{M \times M}$, in which $\mathbf{A}_\ell$ and $\mathbf{B}_\ell $ are defined as
\begin{equation}
\mathbf{A}_\ell \triangleq 
    \begin{cases} 
        \mathbf{W}^{(\ell)} \bm{\Theta}_{\ell-1} \cdots \bm{\Theta}_2 \mathbf{W}^{(2)} \bm{\Theta}_1, & \text{if } \ell \neq 1, \\
        \mathbf{I}_M & \text{if } \ell = 1,
    \end{cases}
\end{equation}
and
\begin{equation}
\mathbf{B}_\ell \triangleq 
    \begin{cases} 
        \bm{\Theta}_L \mathbf{W}^{(L)} \bm{\Theta}_{L-1} \cdots \bm{\Theta}_{\ell+1} \mathbf{W}^{(\ell+1)}, & \text{if } \ell \neq L, \\
        \mathbf{I}_M, & \text{if } \ell = L,
    \end{cases}
\end{equation}
respectively.

\textit{Proof:}  Please, refer to Appendix \ref{ap1}. \hfill $\blacksquare$ \vspace{1mm}

%Proof: See APPENDIX A.1.

\begin{figure*}[t]
    \centering
    \begin{minipage}{0.9\textwidth}    
\begin{algorithm}[H]
\caption{Gradient Descent-Ascent Algorithm.}\label{alg:2}
\begin{algorithmic}[1]
\REQUIRE $\{\mathbf{W}^{(\ell)}\}_{\ell=1}^L$, $\{\mathbf{h}_k\herm\}_{k=1}^K$, $\mathbf{p}$, $\kappa_1$, $\tau$, $\varepsilon$, $\nu_1$, and maximum number of iterations ($I_{max}$).
\STATE \textbf{Initialization:}  Set $i =1$, $\bm{\vartheta}^{(1)} = \mathbf{0}$, $\mu=1$, $\epsilon=1$, and $\bm{\lambda}^{(1)} = \frac{1}{K}\mathbf{1}$.
\FOR{$i=1$ to $I_{max}$}
\STATE According to \eqref{lambda_update}, Update $\bm{\lambda}^{(i+1)}$.
\WHILE{\\$f\left(\bm{\lambda}^{(i+1)},\bm{\vartheta}^{(i)}+\mu\cdot{\nabla_{\bm{\vartheta}}  f\left(\bm{\lambda}^{(i+1)},\bm{\vartheta}^{(i)}\right)}\right)<f\left(\bm{\lambda}^{(i+1)},\bm{\vartheta}^{(i)}\right)+\nu_1\mu{\nabla_{\bm{\vartheta}} f\left(\bm{\lambda}^{(i+1)},\bm{\vartheta}^{(i)}\right)}\tran{\nabla_{\bm{\vartheta}} f\left(\bm{\lambda}^{(i+1)},\bm{\vartheta}^{(i)}\right)}$}
\STATE $\mu= \kappa_1\mu$.
\ENDWHILE
\STATE According to \eqref{psi_update}, update $\bm{\vartheta}^{(i+1)}$.
\STATE Let $\epsilon \leftarrow\tau\mu$.
\STATE \textbf{Until} Convergence  or $i \geq I_{max}$
 \ENDFOR
  \ENSURE Optimal $\bm{\vartheta}$.
\end{algorithmic}
\end{algorithm}
    \end{minipage}
\end{figure*}

 The \ac{GDA} algorithm is summarized in Algorithm \ref{alg:2}. Specifically, when selecting step sizes, we follow the two-time scale \ac{GDA} method explained in \cite{lin2025twotimeGDA}. Thus, we follow that $\epsilon > \mu$ since the minimization step should dominate the maximization step. Therefore, we update $\epsilon= \tau\mu$ where $\tau>0$. After obtaining optimal values for $\bm{\vartheta}$, we finally quantize $\theta_m^{(\ell)}$ into values in $\mathcal{B}$ according to the quantization operation mentioned in \eqref{quantization1}.

The computational complexity of \ac{GDA} algorithm is $\mathcal{O}\left(I_{GDA}\left[LMK^2(4M+3)+K+\log(\varepsilon^{-1})\right]\right)$. Also, the computational complexity of \ac{GP} is $\mathcal{O}(K^{\frac{7}{2}})$ \cite{max_min_RIS_cell_Free}. Thus, total computational complexity is $\mathcal{O}\left(I_{AO}\left(K^{\frac{7}{2}}+I_{GDA}\left[LMK^2(4M+3)+K+\log(\varepsilon^{-1})\right]\right)\right)$, where $I_{GDA}$ and $I_{AO}$ are the iteration need to converge the \ac{GDA} algorithm and alternating optimization algorithm, respectively.  Moreover, the computational complexity of the exhaustive search algorithm is $\mathcal{O}({2^b}^{ML})$, which is significantly higher than the complexity of the proposed algorithm.

\section{JOINT POWER ALLOCATION AND  WAVE-BASED BEAMFORMING WITH THE STATISTICAL CSI}
\label{max-min-erate}
In this section, our main focus is on maximizing the average minimum achievable rate when statistical \ac{CSI} is available at the \ac{SIM}. The advantage of leveraging statistical CSI is the reduction of computational and signaling overhead compared to instantaneous CSI, making SIM optimization more practical for real-world applications. To be specific, the channel covariance matrix $\mathbf{R}_{\text{RIS}}$ is assumed to be known to the \ac{SIM}. Against this backdrop, let us denote the minimum \ac{SINR} among the $K$ users as
\begin{align}
\gamma_{\min}\left(\boldsymbol{\vartheta},\mathbf{p}\right)=\min\left(\gamma_1,\gamma_2,\ldots,\gamma_K\right),
\end{align}
from which we obtain the average minimum achievable rate as
\begin{align}
\label{avgminrate}
R\left(\boldsymbol{\vartheta},\mathbf{p}\right)=\mathbb{E}\left\{\log_2\left(1+\gamma_{\min}\left(\boldsymbol{\vartheta},\mathbf{p}\right)\right)\right\},
\end{align}
where the mathematical expectation is taken with respect to the distribution of $\gamma_{\min}$.
Consequently, we may write the corresponding stochastic optimization problem as
\begin{subequations}
\begin{align}
    (P 4):\ \max _{\mathbf{p}, \boldsymbol{\vartheta}} &\quad \mathbb{E}\left\{\log_2\left(1+\gamma_{\min}\left(\boldsymbol{\vartheta},\mathbf{p}\right)\right)\right\},   \\
    \mbox{subject to} 
      &\quad \sum_{k=1}^{K} p_{k} \leq P_{T}, 
    \\ & \quad p_{k} > 0, \forall k \in \mathcal{K}, 
    \\ &\quad  \theta_m^{(\ell)} \in \mathcal{B},  \forall m \in \mathcal{M}, \forall \ell \in \mathcal{L}.
    %\label{p1_d}
\end{align}
\end{subequations}
The above problem, in its current form, is intractable due to the expectation operator in the objective function. Therefore, to facilitate further analysis, we need to evaluate this quantity. To this end, noting that
\begin{align}
\label{gammak}
    \gamma_k = \frac{|{\mathbf{h}_k\herm} \mathbf{G}_{\bm{\vartheta}}\mathbf{w}_{k}|^2p_k}{\sum_{\substack{j=1 \\ j \neq k}}^K|\mathbf{h}_k\herm \mathbf{G}_{\bm{\vartheta}}\mathbf{w}_j|^2p_j + \sigma^2}
\end{align}
are independent random variables, for $k\in\mathcal{K}$, the \ac{CDF} of $\gamma_{\min}$ can be written as
\begin{align}
    F_{\gamma_{\min}}(z; \boldsymbol{\vartheta},\mathbf{p})&=1-\prod_{k=1}^K \Pr\left\{\gamma_k >z\right\}\nonumber\\
    &=1-\prod_{k=1}^K \left(1-F_{\gamma_k}(z; \boldsymbol{\vartheta},\mathbf{p})\right),
\end{align}
where $F_{\gamma_k}(z;\boldsymbol{\vartheta},\mathbf{p})$ denotes the \ac{CDF} of $\gamma_k$. Therefore, we may use the integration by parts technique to rewrite $\mathbb{E}\left\{\log_2\left(1+\gamma_{\min}\left(\boldsymbol{\vartheta},\mathbf{p}\right)\right)\right\}$ as
\begin{align}
\label{avgrateint}
    &\mathbb{E}\left\{\log_2\left(1+\gamma_{\min}\left(\boldsymbol{\vartheta},\mathbf{p}\right)\right)\right\}\nonumber\\
    & \qquad \qquad =\log_2 e
    \int_0^\infty \frac{\prod_{k=1}^K\left(1-F_{\gamma_k}(z;\boldsymbol{\vartheta},\mathbf{p})\right)}{1+z} {\rm d}z.
\end{align}
Clearly, the main challenge here is to determine the \ac{CDF} of $\gamma_{k}$. This stems from the fact that both the numerator and the denominator of $\gamma_k$ depend on the random vector $\mathbf{h}_k$ and thereby they are correlated. This is further exacerbated by the rank deficiency of the matrix $\sum_{\substack{j=1 \\ j \neq k}}^K \mathbf{w}_j \mathbf{w}_j^H p_j$, which in turn makes  the vector  
$\left(\sum_{\substack{j=1 \\ j \neq k}}^K \mathbf{w}_j \mathbf{w}_j^H p_j\right)^{1/2} \mathbf{G}_{\bm{\vartheta}}^H \mathbf{h}_k$ Gaussian distributed, however with a {\it singular} covariance matrix. Moreover, bounding techniques akin to what is given in \cite{uatf_bound} (i.e., use-and-then-forget bound) are not applicable, since we are not considering large antenna arrays. Therefore, for analytical tractability, we may upper bound $\gamma_k$ as $\gamma_k<\tilde{\gamma}_k$,
where
\begin{align}
   \tilde{\gamma}_k= |{\mathbf{h}_k\herm} \mathbf{G}_{\bm{\vartheta}}\mathbf{w}_{k}|^2\frac{p_k}{ \sigma^2}.
\end{align}
It is also noteworthy that $\tilde{\gamma}_k$ can be interpreted as the \ac{SNR}.

The stochastic inequality $\gamma_k<\tilde{\gamma}_k$ implies $F_{\gamma_k}(z;\boldsymbol{\vartheta},\mathbf{p}) > F_{\tilde{\gamma}_k}(z;\boldsymbol{\vartheta},\mathbf{p}),\; \forall \boldsymbol{\vartheta},\mathbf{p} $.
Therefore, following (\ref{avgrateint}), we conveniently obtain an upper bound on the average minimum rate as
\begin{align}
\label{avgrateintbound}
    &\mathbb{E}\left\{\log_2\left(1+\gamma_{\min}\left(\boldsymbol{\vartheta},\mathbf{p}\right)\right)\right\}\nonumber\\
    & \qquad \qquad =\log_2 e
    \int_0^\infty \frac{\prod_{k=1}^K\left(1-F_{\gamma_k}(z;\boldsymbol{\vartheta},\mathbf{p})\right)}{1+z} {\rm d}z\nonumber\\
    & \qquad \qquad<\log_2 e
    \int_0^\infty \frac{\prod_{k=1}^K\left(1-F_{\tilde{\gamma}_k}(z;\boldsymbol{\vartheta},\mathbf{p})\right)}{1+z} {\rm d}z.
\end{align}
Now noting that ${\mathbf{h}_k\herm} \mathbf{G}_{\bm{\vartheta}}\mathbf{w}_{k}\sim\mathcal{CN}\left(0,\beta_k \mathbf{w}_k^H \mathbf{G}_{\bm{\vartheta}}^H \mathbf{R}_{\text{RIS}}\mathbf{G}_{\bm{\vartheta}} \mathbf{w}_k\right)$, we may write the \ac{CDF} of $\tilde{\gamma}_k$ as
\begin{align}
    F_{\tilde{\gamma}_k}(z;\boldsymbol{\vartheta},\mathbf{p})=1-\exp\left(-\frac{\sigma^2 z}{\beta_kp_k\mathbf{w}_k^H \mathbf{G}_{\bm{\vartheta}}^H \mathbf{R}_{\text{RIS}}\mathbf{G}_{\bm{\vartheta}} \mathbf{w}_k }\right).
\end{align}
Consequently, we obtain from (\ref{avgrateintbound}) 
\begin{align}
\label{avgrateintbound2}
    &\mathbb{E}\left\{\log_2\left(1+\gamma_{\min}\left(\boldsymbol{\vartheta},\mathbf{p}\right)\right)\right\}\nonumber\\
    & \qquad \qquad   <\log_2 e
    \int_0^\infty \frac{\exp\left(-\sigma^2  \zeta \left(\boldsymbol{\vartheta},\mathbf{p}\right)z\right)}{1+z} {\rm d}z,
\end{align}
where 
\begin{align}\label{zeta_func}
\zeta\left(\boldsymbol{\vartheta},\mathbf{p}\right)=\sum_{k=1}^K\frac{1}{\beta_kp_k\mathbf{w}_k^H \mathbf{G}_{\bm{\vartheta}}^H \mathbf{R}_{\text{RIS}}\mathbf{G}_{\bm{\vartheta}} \mathbf{w}_k }.
\end{align}
Alternatively, one can rewrite (\ref{avgrateintbound2}) as
\begin{align}
  \mathbb{E}\left\{\log_2\left(1+\gamma_{\min}\left(\boldsymbol{\vartheta},\mathbf{p}\right)\right)\right\} <\log_2 e\; \mathbb{E}\left\{\frac{1}{\zeta\left(\boldsymbol{\vartheta},\mathbf{p}\right)\sigma^2+z}\right\},
\end{align}
where the expectation is taken with respect to an exponential random variable with the density function $\exp(-z),\; z\geq 0$. Therefore, having motivated with the above bounding process, to facilitate further analysis, we may consider an alternative problem given by
\begin{subequations}
\begin{align}
    (P 5):\ \max _{\mathbf{p}, \boldsymbol{\vartheta}} &\quad \mathbb{E}\left\{\frac{1}{\zeta\left(\boldsymbol{\vartheta},\mathbf{p}\right)\sigma^2+z}\right\} ,   \\
    \mbox{subject to} 
      &\quad \sum_{k=1}^{K} p_{k} \leq P_{T}, 
    \\ & \quad p_{k} > 0, \forall k \in \mathcal{K}, 
    \\ &\quad  \theta_m^{(\ell)} \in \mathcal{B},  \forall m \in \mathcal{M}, \forall \ell \in \mathcal{L}.
\end{align}
\end{subequations}
A careful inspection of the above problem reveals that it is an alternative formulation corresponding to the maximization of the average minimum achievable rate given by
\begin{align}
    \tilde{R}\left(\boldsymbol{\vartheta},\mathbf{p}\right) =\mathbb{E}\left\{\log_2\left(1+\tilde{\gamma}_{\min}\left(\boldsymbol{\vartheta},\mathbf{p}\right)\right)\right\},
\end{align}
where $\tilde{\gamma}_{\min}\left(\boldsymbol{\vartheta},\mathbf{p}\right)=\min\left(\tilde{\gamma}_1,\tilde{\gamma}_2,\ldots,\tilde{\gamma}_K\right)$. Here, the notable difference is that we consider the \ac{SNR} instead of the \ac{SINR}. This approach, however with the instantaneous \ac{CSI}, has been adopted in \cite{max_min_RIS_GDA}. Therefore, we obtain an optimistic average minimum rate in comparison with
the case corresponding to the \ac{SINR}. Nevertheless, as our simulation results reveal, this bound serves as a good approximation for the true average rate in the low \ac{SNR} regime.   The utility of this approach should be viewed in conjunction with the extreme analytical complexity associated with the original problem involving the \ac{SINR}. Furthermore, our proposed approach reduces the complexity of the original optimization since it does not require updating the policy in every coherence interval. 

Now, keeping in mind that $\mathbb{E}\left\{\frac{1}{\zeta\left(\boldsymbol{\vartheta},\mathbf{p}\right)\sigma^2+z}\right\}$ is convex and decreasing in $\zeta\left(\boldsymbol{\vartheta},\mathbf{p}\right)$ \cite{boyd2004convex}, the problem (\emph{P}5) can equivalently be written as
\begin{subequations}
\begin{align}
    (P 6):\ \min _{\mathbf{p}, \boldsymbol{\vartheta}} &\quad \sum_{k=1}^K\frac{1}{\beta_kp_k\mathbf{w}_k^H \mathbf{G}_{\bm{\vartheta}}^H \mathbf{R}_{\text{RIS}}\mathbf{G}_{\bm{\vartheta}} \mathbf{w}_k } ,   \\
    \mbox{subject to} 
      &\quad \sum_{k=1}^{K} p_{k} \leq P_{T}, \label{p6_b}
    \\ & \quad p_{k} > 0, \forall k \in \mathcal{K}, \label{p6_c}
    \\ &\quad  \theta_m^{(\ell)} \in \mathcal{B},  \forall m \in \mathcal{M}, \forall \ell \in \mathcal{L} \label{p6_d}.
\end{align}
\end{subequations}
The optimal policies corresponding to the above problem (P6), given by $\mathbf{p}=\mathbf{\tilde p}^\star$ and $\bm{\vartheta}=\bm{\tilde\vartheta}^\star$, yield the maximum of the minimum achievable rate
\begin{align}
    \tilde{R}\left(\boldsymbol{\tilde\vartheta}^\star,\mathbf{\tilde p}^\star\right)&=\log_2 e\; \mathbb{E}\left\{\frac{1}{\zeta\left(\boldsymbol{\tilde\vartheta}^\star,\mathbf{\tilde p}^\star\right)\sigma^2+z}\right\}\nonumber\\
&=\exp\left(\zeta\left(\boldsymbol{\tilde\vartheta}^\star,\mathbf{\tilde p}^\star\right)\sigma^2\right)E_1\left(\zeta\left(\boldsymbol{\tilde \vartheta}^\star,\mathbf{\tilde p}^\star\right)\sigma^2\right),
\end{align}
where $E_1(z)=\int_z^\infty \frac{\exp(-t)}{t} {\rm d}t,\; z>0$ is the exponential function. Consequently, we can relate the above rate to the desired optimal rate corresponding to the original stochastic optimization problem (\emph{P}4) as 
\begin{align}
R\left(\boldsymbol{\vartheta}^\star,\mathbf{p}^\star\right)< \tilde{R}\left(\boldsymbol{\tilde \vartheta}^\star,\mathbf{\tilde p}^\star\right),
\end{align}
where $\mathbf{p}=\mathbf{p}^\star$ and $\boldsymbol{\vartheta}=\boldsymbol{\vartheta}^\star$ are the optimal policies for the original  problem (\emph{P}4).

Having been motivated by the above reasoning, we now focus on determining the optimal policies for the problems (\emph{P}5) and (\emph{P}6). We propose an alternating optimization algorithm to solve the problem (\emph{P}6). Specifically, we solve the power allocation problem in the first sub-problem and the wave-based beamforming problem in the second sub-problem, as listed below. We solve these two sub-problems iteratively until convergence to find $\mathbf{\tilde p}^\star$ and $\bm{\tilde\vartheta}^\star$ similarly as the Algorithm \ref{alg:1}. 
\begin{itemize}
    \item  The power allocation optimization problem for fixed \ac{SIM} phase shifts can be formulated as
\begin{subequations}
\begin{align}
    (P 6.1):\ \min _{\mathbf{p}} &\quad \sum_{k=1}^K\frac{1}{\beta_kp_k\mathbf{w}_k^H \mathbf{G}_{\bm{\vartheta}}^H \mathbf{R}_{\text{RIS}}\mathbf{G}_{\bm{\vartheta}} \mathbf{w}_k } ,   \\
    \mbox{subject to} 
      &\quad  \mbox{constraints~} \eqref{p6_b}, \eqref{p6_c}.
\end{align}
\end{subequations}

Thus, the objective function and the constraints are monomial and polynomial functions. Hence,
 the optimization problem (\emph{P}6.1) follows the standard form of the \ac{GP} problem \cite{boyd2004convex}, and it can be solved using standard optimization tools.
 \item The optimal wave-based beamforming problem for a given power allocation can be formulated as 
\begin{subequations}
\begin{align}
    (P 6.2):\ \min _{ \boldsymbol{\vartheta}} &\quad \sum_{k=1}^K\frac{1}{\beta_{k} p_{k}\mathbf{w}_k^H \mathbf{G}_{\bm{\vartheta}}^H \mathbf{R}_{\text{RIS}}\mathbf{G}_{\bm{\vartheta}} \mathbf{w}_k } ,   \\
    \mbox{subject to} 
      &\quad  \mbox{constraints~}  \eqref{p6_d}.
\end{align}
\end{subequations}

We propose a gradient descent-based algorithm to solve the wave-based beamforming optimization problem. Thus, all the \ac{SIM} phase shifts $\bm{\vartheta}$ can be updated using gradient descent step as follows:
\begin{align}
    {{\bm{\vartheta}}}^{(i+1)} &=  {\bm{\vartheta}}^{(i)} -  \iota\cdot{\nabla_{\bm{\vartheta}}  \zeta(\mathbf{p},\bm{\vartheta}^{(i)})}. \label{psi_update2}
 \end{align}
 Here, $\iota$ is the step size which is updated using a backtracking line search \cite{boyd2004convex}. 
Specifically, ${\nabla_{\bm{\vartheta}} \zeta} = \left[ \frac{\partial \zeta}{\partial {\theta_1^{(1)}}} ,\cdots, \frac{\partial \zeta}{\partial {\theta_M^{(1)}}}, \cdots,  \frac{\partial \zeta}{\partial {\theta_1^{(L)}}}, \cdots,  \frac{\partial \zeta}{\partial {\theta_M^{(L)}}}\right]\tran$  is the gradient of $\zeta$ with respect to $\bm{\vartheta}$, where $  \frac{\partial \zeta}{\partial {\theta_m^{(\ell)}}}$ is obtained as follows. \vspace{1mm}

\end{itemize}

\paragraph*{Lemma II}\label{lemma:gradient2}
The gradient of the objective function $\zeta$, defined in \eqref{zeta_func}, with respect to the phase shift $\theta_m^{(\ell)}$ at the $m$-th element of the $\ell$-th metasurface layer is given by:
\begin{equation}
    \frac{\partial \zeta}{\partial {\theta_m^{(\ell)}}} = \sum_{k=1}^K    \frac{1}{\beta_k p_k} \frac{- c_{m,k}^{(\ell)} }{\left (\sum_{i=1}^M \rho_i|\bm{\upsilon}_i\herm\mathbf{G}_{\bm{\vartheta}}\mathbf{w}_{k}|^2 \right)^2}, 
\end{equation}
where \begin{equation}
    c_{m,k}^{(\ell)} = 2\sum_{i=1}^M \rho_i \Im\left[ e^{-\mathsf{j}\theta_m^{(\ell)}}(\mathbf{w}_k)\herm\mathbf{a}^\ell_m(\mathbf{b}^\ell_m)\herm\bm{\upsilon}_i\bm{\upsilon}_i\herm \mathbf{G}_{\bm{\vartheta}}\mathbf{w}_k\right],
\end{equation}
with $\rho_i$ and $\bm{\upsilon}_i$ denoting, respectively, the eigenvalues and eigenvectors of $\mathbf{R}_{\text{RIS}}$.

\textit{Proof:} Please refer to Appendix \ref{ap2}. \hfill $\blacksquare$ \vspace{1mm}

The gradient descent algorithm to find optimal $\bm{\vartheta}$ is described in Algorithm \ref{alg:3}.
\begin{algorithm}[t]
\caption{Gradient Descent Algorithm.}\label{alg:3}
\begin{algorithmic}[1]
\REQUIRE $\{\mathbf{W}^{(\ell)}\}_{\ell=1}^L$, $\mathbf{R}_{\text{RIS}}$,  $\mathbf{p}$, $\kappa_2$, $\nu_2$, and $I_{max}$.
\STATE \textbf{Initialization:}  Set $i =1$, $\bm{\vartheta}^{(1)} = \mathbf{0}$, $\iota=1$.
\FOR{$i=1$ to $I_{max}$}
\WHILE{$\zeta\left(\mathbf{P},\bm{\vartheta}^{(i)} -\iota\cdot{\nabla_{\bm{\vartheta}}  \zeta\left(\mathbf{P},\bm{\vartheta}^{(i)}\right)}\right) > \zeta\left(\mathbf{P},\bm{\vartheta}^{(i)}\right) - \nu_2\iota{\nabla_{\bm{\vartheta}}  \zeta\left(\mathbf{P},\bm{\vartheta}^{(i)}\right)}\tran{\nabla_{\bm{\vartheta}}  \zeta\left(\mathbf{P},\bm{\vartheta}^{(i)}\right)}$}
\STATE $\iota= \kappa_2\iota$
\ENDWHILE
\STATE According to \eqref{psi_update2}, update $\bm{\vartheta}^{(i+1)}$.
\STATE \textbf{Until} Convergence  or $i \geq I_{max}$
 \ENDFOR
  \ENSURE Optimal $\bm{\vartheta}$.
\end{algorithmic}
\end{algorithm}
After obtaining optimal values for $\bm{\tilde\vartheta}^\star$, we finally quantize $\theta_m^{(\ell)}$ into values in $\mathcal{B}$ using the quantization operation in \ref{quantization1}. 
The total computational complexity of the alternating algorithm is $\mathcal{O}\left(I_{AO}\left(K^{\frac{7}{2}}+I_{GD}\left[LMK(4M^2+2M+1)\right]\right)\right)$, where $I_{GD}$ and $I_{AO}$ are the iterations need to converge the gradient descent algorithm and alternating optimization algorithm, respectively. 
%Now, a careful inspection of $\gamma_k$ in (\ref{gammak}) reveals that its constituent random variables $|{\mathbf{h}_k\herm} \mathbf{G}_{\bm{\vartheta}}\mathbf{w}_{k}|^2$ and $\sum_{\substack{j=1 \\ j \neq k}}^K|\mathbf{h}_k\herm \mathbf{G}_{\bm{\vartheta}}\mathbf{w}_j|^2p_j$ are correlated. This in turn makes it challenging to evaluate the exact \ac{CDF} of $\gamma_k$. Therefore, for conveneince, let us represent $\gamma_k$ as
%\begin{align}
  %  \gamma_k=
%\end{align}

\section{NUMERICAL RESULTS}
\label{results}
\begin{figure}[t]
\centering
   \includegraphics[width=0.95\columnwidth]{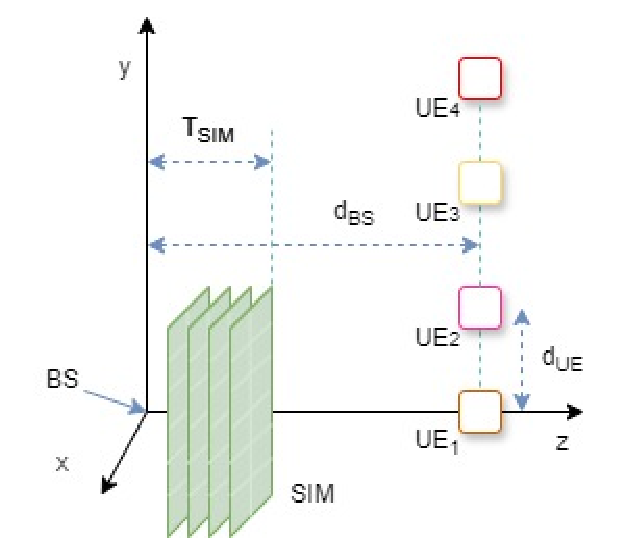 }
  \caption{Illustration of the simulation setup.
  \label{fig:Illustration of the simulation setup}}
\end{figure}

In this section, we provide simulation results to evaluate the performance of the proposed max-min \ac{SINR} algorithm. Here, we consider a downlink of a \ac{SIM}-assisted multi-user \ac{MIMO} system. Fig. \ref{fig:Illustration of the simulation setup} shows an illustration of the simulation setup. The \ac{BS} has an $M$-antenna array along the x-axis, centred at  $(0,0,0)$. Each \ac{SIM} layer is positioned parallel to the x-y plane, and each layer's centre is aligned with the z-axis. The first \ac{SIM} layer is spaced with $\rm{T}_{\rm{SIM}}$$/L$ distance from the centre of the \ac{BS} and the rest of the layers are spaced with $\rm{T}_{\rm{SIM}}$$/L$  distance from each other, where $\rm{T}_{\rm{SIM}}$ is the thickness of the \ac{SIM}. \ac{UE}s are located parallel to the y-axis, with perpendicular distance $\rm{d}_{\rm{BS}}$ from the \ac{BS} and $\rm{d}_{\rm{UE}}$ distance between each \ac{UE}.
The distance-dependent path loss is modeled as  
\begin{equation}
    \beta_k = C_0(\rm{d}_k/\rm{d}_0)^{(-\alpha)}, \rm{d}_k \geq \rm{d}_0,
\end{equation}
where $\rm{d}_k = \sqrt{(\rm{d}_{\rm{BS}}- T_{\rm{SIM}})^2 + (\rm{d}_{\rm{UE}}(k-1))^2}$ is the distance between \ac{SIM} and $k$-th \ac{UE}. Additionally,  $5$~dBi antenna gain is present at each antenna of the \ac{BS}. Table \ref{table:simulation parameters} lists the rest of the simulation parameters.
\begin{table}[t]
	\caption{Simulation Parameters} % title of Table
	\centering % used for centering table
	\begin{tabular}{|l| l| } 
		\hline 
        Carrier frequency, $f_c$ & $28$~GHz \\
        \hline 
        Length and width of each \ac{RIS} element , $\rm{d}_x, \rm{d}_y$  & $\lambda/2$\\
        \hline
        Thickness of \ac{SIM}, $\rm{T}_{\rm{SIM}}$,& $5\lambda$ \\
        \hline 
        The distance between \ac{UE}s,  $\rm{d}_{\rm{UE}}$& $10$~m \\
        \hline
        The perpendicular distance between \ac{BS} and \ac{UE} , $\rm{d}_{\rm{BS}}$ & $10$~m \\ 
        \hline
        Free space path loss, $C_0$ &$-30$~dB \\
        \hline
        The path loss exponent $\alpha$ & $3.5$ \\
        \hline
        The reference distance, $\rm{d}_0$&$1$~m \\
        \hline
        Noise variance for each \ac{UE}, $\sigma_k^2$ &$-90$~dBm \\
        \hline
	\end{tabular}
	\label{table:simulation parameters} 
	%\vspace{-2mm}
\end{table}

\subsection{JOINT POWER ALLOCATION AND WAVE-BASED BEAMFORMING WITH  THE INSTANTANEOUS CSI}
\label{I_CSI_result}
To the best of our knowledge, no other works have investigated a fairness-based resource allocation for a SIM-aided system. Thus, we compare the performance of the proposed max-min rate optimization algorithm based on instantaneous \ac{CSI} with the standard approaches listed below.
%The proposed max-min rate optimization algorithm with instantaneous \ac{CSI} and benchmark schemes we used to compare our proposed algorithm are listed below:
\begin{itemize}
    \item \ac{GP}+GDA(Continuous phase shift): We solve the power allocation problem with \ac{GP} and wave-based beamforming optimization algorithm for continuous phase shifts.
    \item Proposed: (GP+GDA): The proposed max-min rate algorithm with instantaneous \ac{CSI} using \ac{GP} for power allocation and \ac{GDA} algorithm with quantization for wave-based beamforming. Here, we use the parameters:  $\kappa_1=0.8$, $\nu_1=0.3$, $\tau=10$,  $\varepsilon=10^{-4}$, and $b=8$.
    \item Equal power + \ac{GDA}: We use equal power allocation between \ac{UE}s by dividing the total power equally among \ac{UE}s and the proposed \ac{GDA} algorithm for wave-based beamforming optimization.
    \item \ac{GP} + random phase shifts:  We use optimal power allocation found using \ac{GP} and randomly selected phase shift in $\mathcal{B}$.    
    \item Equal power  +random phase shift: We use equal power allocation between \ac{UE}s and randomly selected phase shift in $\mathcal{B}$.    
    %\item Exhaustive search:  We solve the power allocation problem with \ac{GP} programming and wave-based beamforming optimization with an exhaustive search algorithm when $b=1$. 
\end{itemize}

\begin{figure}[t]
    \centering
    \includegraphics[width=1\columnwidth]{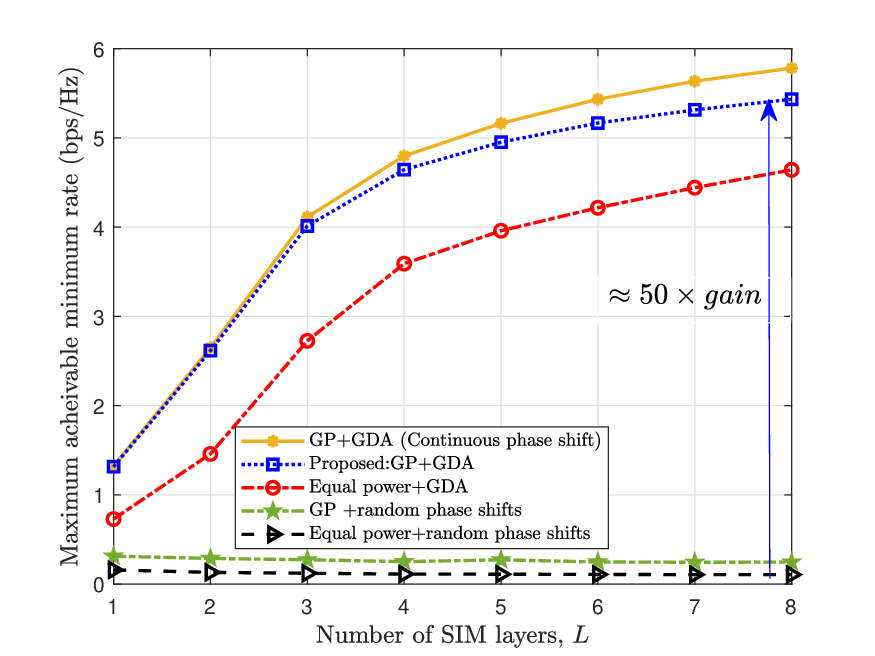}
     \caption{The maximum achievable minimum data rate versus the number of \ac{SIM} layers, $L$. The simulation is conducted with the parameters $N=K=4$, $M=36$, and $P_T=10$~dBm.}
     \label{Fig:L} \vspace{-3mm}
\end{figure}
Fig. \ref{Fig:L}  shows the maximum achievable minimum rate variation with the number of \ac{SIM} layers. It shows that the minimum rate increases with the number of \ac{SIM} layers. Also, our proposed optimization scheme gives an approximate $50$ times higher minimum rate than equal power allocation and random phase shift assignment. Further, our proposed algorithm can achieve an approximate $1.2$ times higher minimum rate compared to equal power allocation with optimal phase shift assignment and an approximate $20$ times higher minimum rate compared to optimal power allocation and random phase shift assignment when $L=8$.  Moreover, the minimum rate increases fourfold when compared to using one \ac{SIM} layer with eight \ac{SIM} layers, which proves the performance improvement capability of using \ac{SIM} with multiple \ac{RIS} layers. Additionally, the gap between our proposed optimization algorithm with quantization and continuous phase shift increases with the increase of \ac{SIM} layers since the effect of quantization error increases with the increase in the number of phases that need to be optimized. However, the highest rate loss we observed due to the quantization is approximately $0.35$~bps when $L=8$.

\begin{figure}[t]
    \centering
    \includegraphics[width=1\columnwidth]{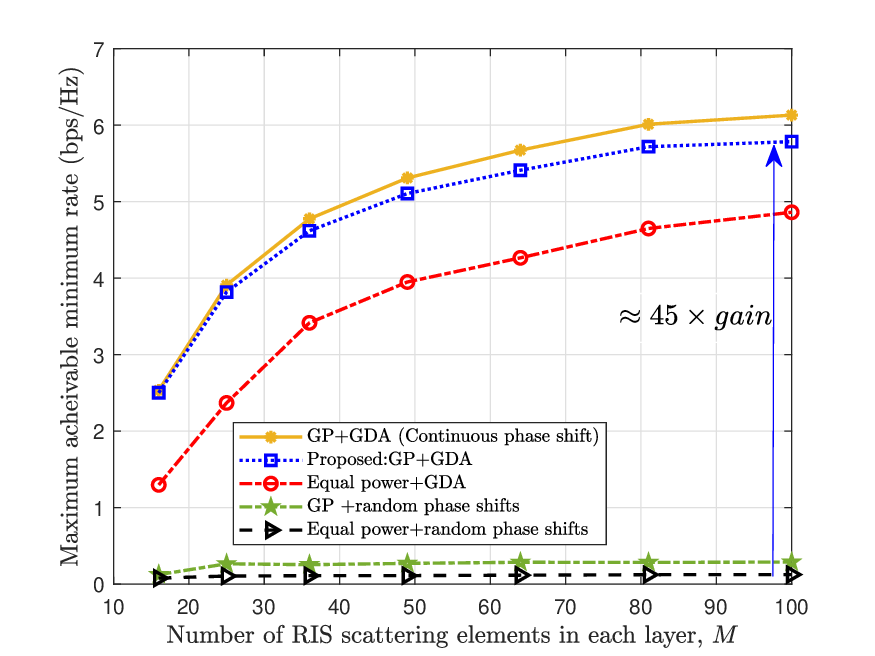}
     \caption{The maximum achievable minimum data rate versus the number of scattering elements in each layer, $M$. The simulation is conducted with the parameters $N=K=4$, $L=4$, and $P_T=10$~dBm.}
     \label{Fig:M} \vspace{-4mm}
\end{figure}
The maximum achievable minimum rate variation with the number of \ac{RIS} elements in each layer is shown in Fig. \ref{Fig:M}. It shows that the minimum rate increases with the number of \ac{RIS} elements. Also, our proposed optimization scheme gives an approximate $45$ times higher minimum rate than equal power allocation and random phase shift assignment. Further, our proposed algorithm can achieve an approximate $1.2$ times higher minimum rate compared to equal power allocation with optimal phase shift assignment and an approximate $20$ times higher minimum rate compared to optimal power allocation and random phase shift assignment when $M=100$. Moreover, the gap between our proposed optimization algorithm with quantization and continuous phase shift increases with the increase of \ac{RIS} elements since the effect of quantization error increases with the increase in the number of phases that need to be optimized.  However, the highest rate loss we observed due to the quantization is approximately $0.35$~bps when $M=100$.

%Fig. \ref{fig:PT} and \ref{fig:PTEX}  show the maximum achievable minimum rate variation with different total transmit power values for L = 2 and L = 4, respectively.  They show that increased transmit power at the \ac{BS} increases the minimum rate. Also, our proposed optimization scheme gives a 64 times higher minimum rate than equal power allocation and random phase shift assignment. Further, our proposed algorithm can achieve a 1.3 times higher minimum rate compared to equal power allocation with optimal phase shift assignment and a 26 times higher minimum rate compared to optimal power allocation and random phase shift assignment when the total power at \ac{BS} is $30~dBm$ and $M=36$, $L=4$. Moreover, the highest rate loss we observed due to the quantization is $0.1~bps$. The computational complexity is growing exponentially with $M$, $L$ and $b$. To show the comparison with exhaustive search with the proposed algorithm, we have added Fig \ref{fig:PTEX} for low values of $M$ and $L$. According to  Fig. \ref{fig:PTEX},  our proposed algorithm achieves a higher minimum rate than the exhaustive search algorithm, since we use 8 bits for quantization in our proposed method and only 1 bit in the exhaustive search. Even though it is not a fair comparison, this hints that our proposed method gives optimal solutions.
\begin{figure}[t]
    \centering
	\includegraphics[width=1\columnwidth]{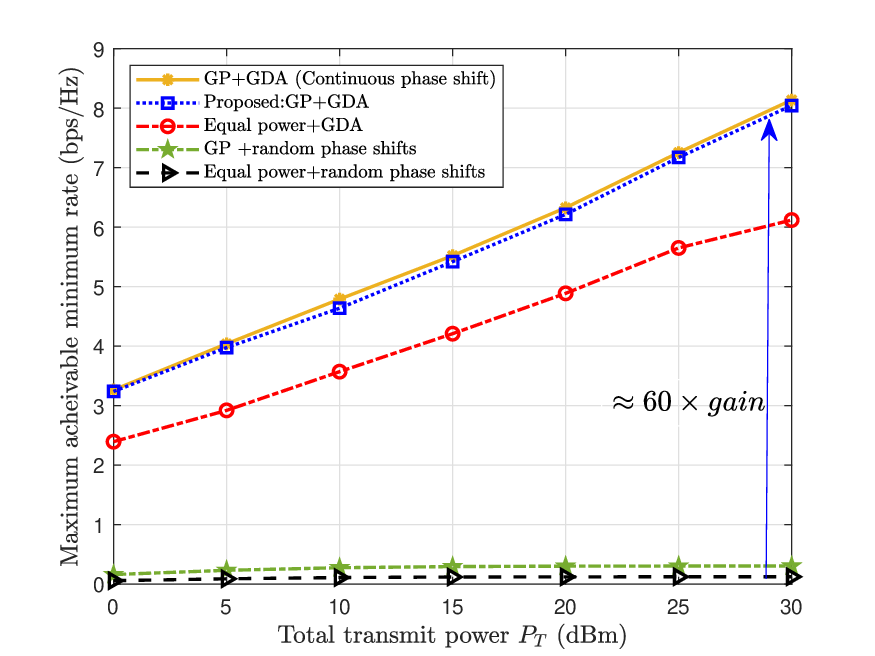}
     \caption{The maximum achievable minimum data rate versus the total transmit power, $P_T$. The simulation is conducted with the parameters $N=K=4$, $M=36$, $L=4$. }	
     \label{fig:PT} \vspace{-4mm}
\end{figure}
Fig. \ref{fig:PT}  shows the maximum achievable minimum rate variation with different total transmit power values for $L = 4$.  They show that increased transmit power at the \ac{BS} increases the minimum rate. Also, our proposed optimization scheme gives an approximate $60$ times higher minimum rate than equal power allocation and random phase shift assignment. Further, our proposed algorithm can achieve an approximate $1.3$ times higher minimum rate compared to equal power allocation with optimal phase shift assignment and an approximate $25$ times higher minimum rate compared to optimal power allocation and random phase shift assignment when the total power at \ac{BS} is $30$~dBm and $M=36$, $L=4$. Moreover, the highest rate loss we observed due to the quantization is $0.1$~bps. 

\begin{figure}[t] 
    \centering
    \includegraphics[width=1\columnwidth]{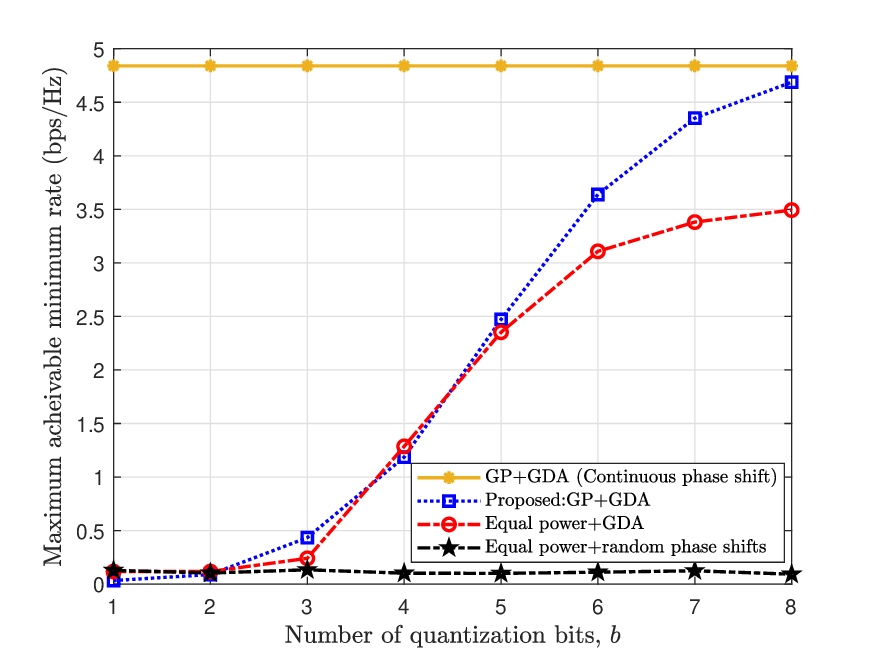}
     \caption{The maximum achievable minimum data rate versus the number of quantization bits, $b$. The simulation is conducted with the parameters $N=K=4$, $M=36$, $L=4$, and  $P_T=10$~dBm.}
      \label{Fig:Bit} 
\end{figure}
Fig. \ref{Fig:Bit} shows the performance variation of the algorithm for different numbers of bits used to quantize the phase shift. In Fig. \ref{Fig:Bit}, it is visible that when $b=8$, our proposed algorithm achieved the performance of the \ac{GDA} algorithm when we assume continuous phase shifts. Only a $0.1$~bps rate loss occurs due to the quantization using 8 bits. However, due to the quantization error, the rate loss is high if $b$ is 1 to 5.

\begin{figure}[t]
    \centering
    \includegraphics[width=1\columnwidth]{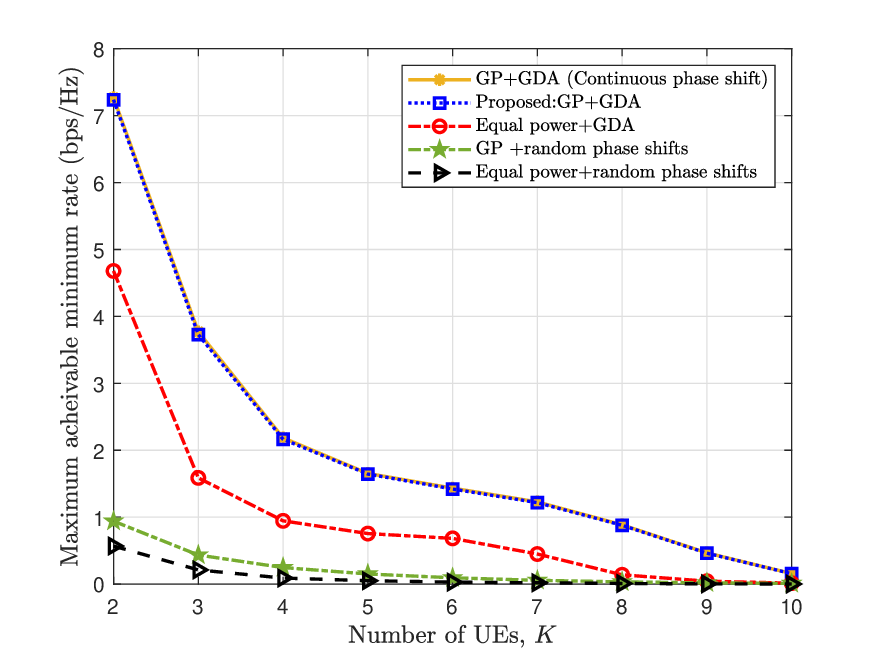}
     \caption{The maximum achievable minimum data rate versus the number of \ac{UE}s, $K$. The simulation is conducted with the parameters $N=10$, $M=36$, $L=4$, and $P_T=10$~dBm.}
      \label{Fig:K} 
\end{figure}
The maximum achievable minimum rate variation with the number of \ac{UE}s is shown in Fig. \ref{Fig:K}. According to Fig. \ref{Fig:K}, the minimum rate decreases as the number of \ac{UE}s grows since the limitation in power at the \ac{BS}. Moreover, when the system has a very high number of \ac{UE}s, i.e., $K>5$, the system gives a very low minimum rate for the weaker \ac{UE}. However,  our proposed optimization scheme gives an approximate $15$ times higher minimum rate than equal power allocation and random phase shift assignment, even for $10$ \ac{UE}s.

All the simulations in Section \ref{results}-\ref{I_CSI_result} are carried out for 1000 channel realizations. In summary, the proposed alternating optimization algorithm performs better with benchmark schemes: equal power allocation with \ac{GDA}, \ac{GP} power allocation with random phase shifts, equal power allocation with random phase shifts, and \ac{GP} power allocation with exhaustive search. In addition, the proposed algorithm performs almost similarly to the \ac{GDA} with continuous phase shift values when we quantize using 8 bits. Moreover, it is visible that finding optimal \ac{SIM} phase shifts contributes more to improving performance than optimal power allocation.

\begin{figure}[t]
    \centering
    \includegraphics[width=1\columnwidth]{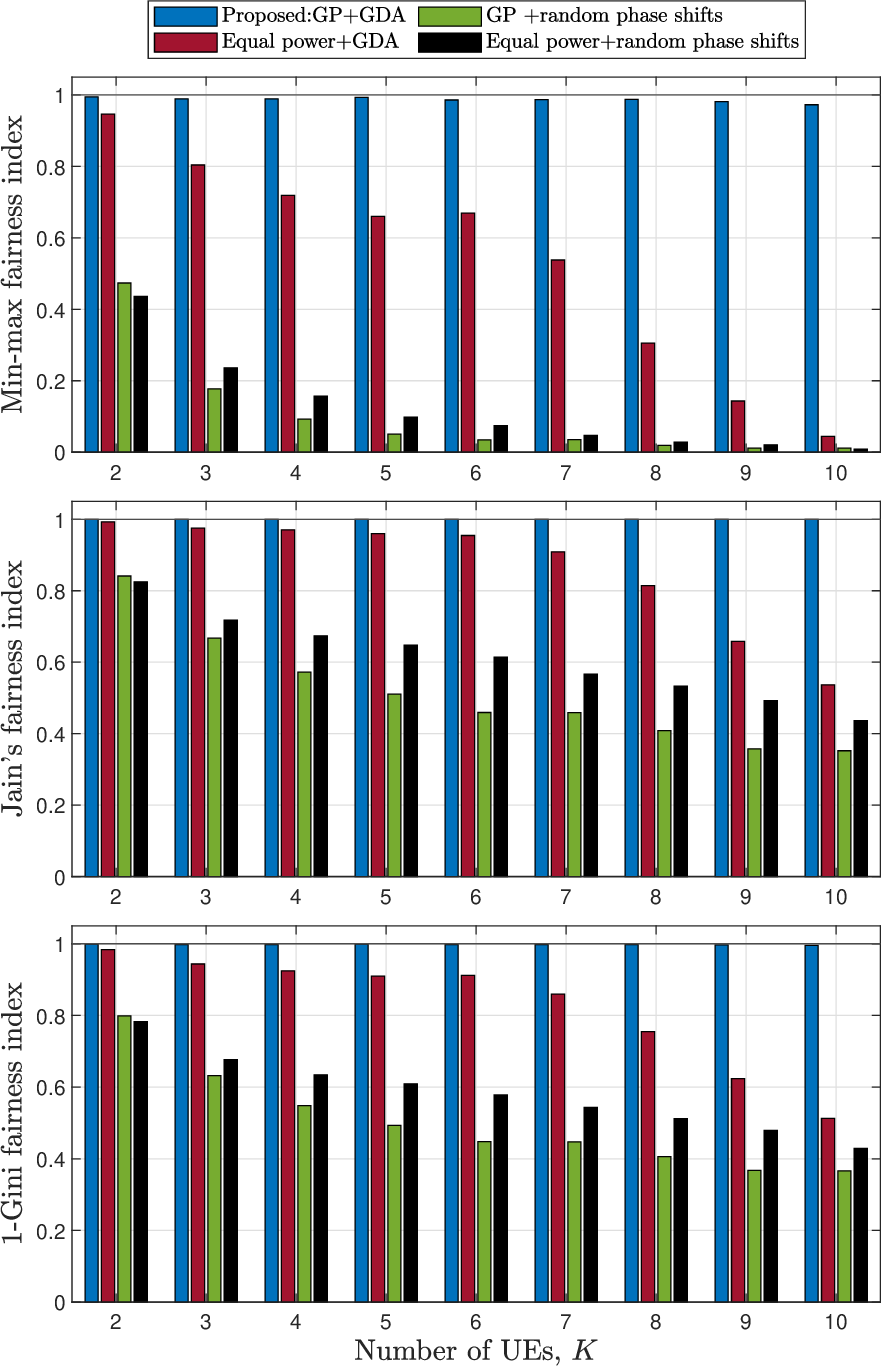}
     \caption{The fairness index versus the number of \ac{UE}s, $K$. The simulation is conducted with the parameters $N=10$, $M=36$, $L=4$, and $P_T=10~dBm$.}
      \label{Fig:fairness_index}
\end{figure}
To further investigate our proposed algorithm's performance, we analyze its resource allocation's fairness. Specifically, we utilize the following three fairness indices, which are frequently used in wireless communication systems  \cite{FI_intro1,FI_intro2,FI_intro3,FI_intro4} to analyze the fairness of our proposed methods as well as the used benchmark schemes.
\begin{align*}
    &\text{Min-max  fairness index} \hspace{1cm} I_{\min-\max}=\frac{\min{(R_k)}}{\max{(R_k)}}, \\
    &\text{Jain's fairness index}\hspace{1.5cm}  I_{\text{Jain}}=\frac{|\sum_{k=1}^KR_k|^2}{K\sum_{k=1}^KR_k^2}, \\
    &\text {Gini fairness index} \hspace{0.2cm}  I_{\text{Gini}}=\frac{1}{2K^2\bar{R}}\sum_{i=1}^K\sum_{j=1}^K |R_i-R_j|,
\end{align*}
where $\bar{R}=(1/K)\sum_{j=1}^K R_j$ is the average rate. All three fairness indexes vary from $1$ to $0$. Both the min-max fairness and Jain's fairness indices result in a value of $1$, indicating a high degree of fairness, whereas a value of $0$ indicates complete unfairness.  In contrast, the Gini fairness index results in a value of $0$, indicating a high degree of fairness, whereas a value of $1$ indicates complete unfairness. Therefore, here we consider $1-I_{\text{Gini}}$ for ease of comparison.  Fig. \ref{Fig:fairness_index} shows the variation of the above-mentioned three fairness indices with the number of \ac{UE}s for our proposed 'GP+GDA' max-min fairness algorithm and the benchmark schemes.  According to Fig. \ref{Fig:fairness_index}, the proposed max-min fairness algorithm exhibits that the fairness indices are approximately equal to $1$ even when the system has 10 \ac{UE}s. This demonstrates that our proposed algorithm exhibits high fairness irrespective of the number of users. In contrast, the benchmark schemes show a reduction in fairness when the number of \ac{UE}s grow.  Thus, this proves that our proposed algorithm can allocate resources such that the system is highly fair for every \ac{UE}.

\begin{figure}[t]
\centering
\includegraphics[width=1\columnwidth]{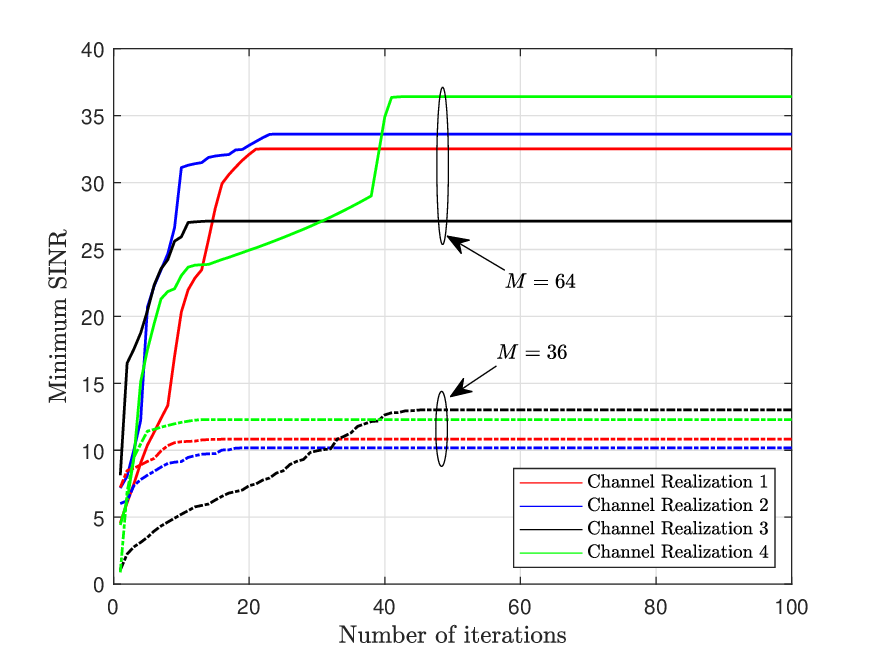}
\caption{Convergence behavior of max-min fairness alternating optimization algorithm with instantaneous \ac{CSI}. }	\label{fig:convergence}\vspace{-3mm}
\end{figure}
Fig. \ref{fig:convergence} shows the convergence behavior of max-min fairness optimization algorithm with instantaneous \ac{CSI} for the alternating optimization algorithm. It shows that the proposed algorithm converges in a reasonable number of iterations. 

\subsection{JOINT POWER ALLOCATION AND  WAVE-BASED BEAMFORMING WITH THE STATISTICAL CSI}

To the best of our knowledge, no other works have investigated a fairness-based resource allocation for a SIM-aided system. Thus, we compare the performance of the proposed max-min rate optimization algorithm based on statistical \ac{CSI} with the standard approaches listed below.
%The proposed max-min average rate optimization algorithm with statistical \ac{CSI} and benchmark schemes we used to compare our proposed algorithm are listed below:

\begin{itemize}
    \item Proposed: Upper bound: The proposed algorithm to maximize the average minimum achievable rate with statistical \ac{CSI} using \ac{GP} programming for optimal power allocation and gradient descent algorithm with quantization for wave-based beamforming. Here, we use the parameters:  $\kappa_2=0.8$, $\nu_2=0.3$, and $b=3$.   
    \item Exhaustive search:  We solve the optimization problem (\emph{P}4) with an exhaustive search algorithm when $b=1$. Here, we assume total transmit power, $P_T$ is a discrete variable with $0.01$ resolution.
\end{itemize}

\begin{figure}[t]
    \centering
    \includegraphics[width=1\columnwidth]{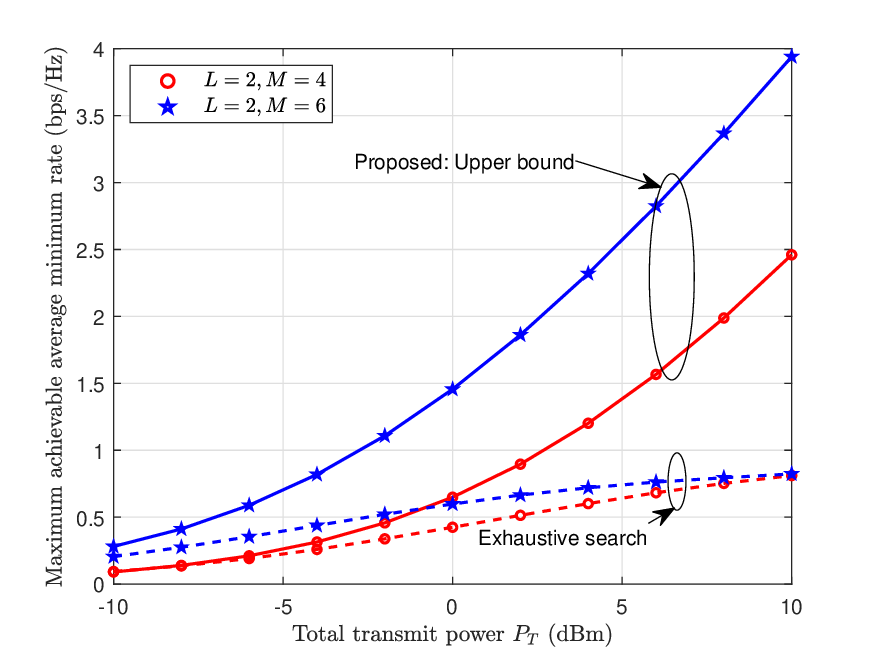}     
      \caption{The maximum achievable average minimum rate versus the total transmit power, $P_T$.  The simulation is conducted with the parameters $N=K=2$.}
      \label{Fig:PT_ECEX}\vspace{-3mm}
\end{figure}
Fig. \ref{Fig:PT_ECEX} compares our proposed upper bound and the results obtained exhaustively for the maximum achievable average minimum rate. Further,  it shows the variation of the maximum achievable average minimum rate with transmit power. According to Fig. \ref{Fig:PT_ECEX}, it exhibits a tight upper bound for low transmit power at \ac{BS}, in other words, during the low \ac{SNR} regime. (Since we assume noise power is constant throughout this simulation). Thus, it proves the applicability of the proposed upper bound even if it is overly optimized in the high \ac{SNR} regime. Therefore, we are able to reduce the frequency of finding optimal power allocation and wave-based beamforming using the proposed upper bound for statistical \ac{CSI}. We have done this simulation for low $L$ and $M$ values due to the high complexity of the exhaustive algorithm. 

\begin{figure}[t]
    \centering
    \includegraphics[width=1\columnwidth]{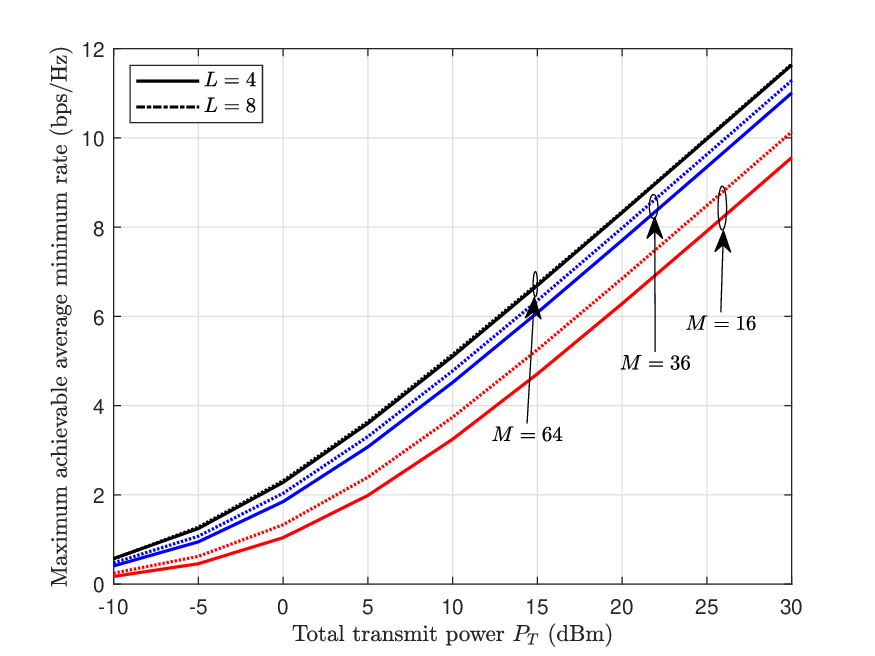}
      \caption{The proposed upper bound for the maximum achievable average minimum rate versus the total transmit power at \ac{BS}, $P_T$. The simulation is conducted with the parameters $N=K=4$.}
       \label{Fig:PT_EC} \vspace{-3mm}
\end{figure}
The variation of the upper bound for the maximum achievable average minimum rate with transmit power and number of \ac{RIS} elements at each \ac{SIM} layer and number of \ac{SIM} layers is shown in Fig. \ref{Fig:PT_EC}. % Moreover, the variation of the upper bound for the maximum achievable average minimum rate with a number of \ac{SIM} layers and a number of \ac{RIS} elements at each \ac{SIM} layer is shown in Fig. \ref{fig:L_EC} for the low \ac{SNR} regime, in which our proposed upper bound works perfectly.
According to Fig. \ref{Fig:PT_EC}, it exhibits an increase in the upper bound with the increase of $P_T$, $L$ and $M$.

\begin{comment}
 \begin{figure}[t]
    \centering
    \includegraphics[width=0.9\columnwidth]{results_v8_3/L_loop_upperbound_0dbm.eps}
        \caption{The proposed upper bound for the maximum achievable average minimum rate versus the number of \ac{SIM} layers, $L$. The simulation is conducted with the parameters $N=K=4$ and $P_T=0~dBm$.}
         \label{fig:L_EC}
\end{figure}
\end{comment}
\begin{figure}[t]
    \centering      
		\includegraphics[width=1\columnwidth]{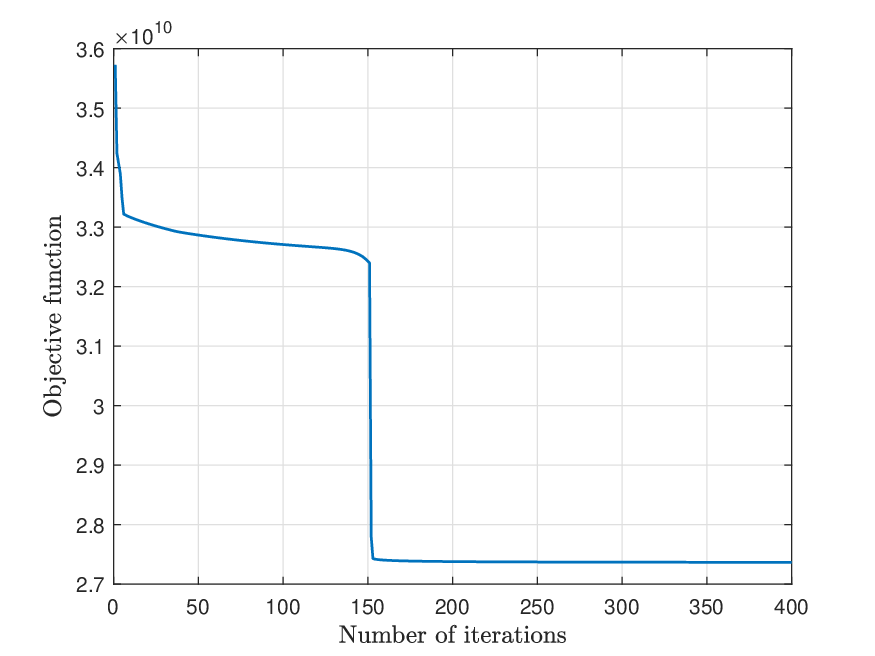}
 \caption{Convergence behavior of max-min fairness alternating optimization algorithm with statistical \ac{CSI}. }
	\label{fig:convergenceEC}
\end{figure}
Fig. \ref{fig:convergenceEC} shows the convergence behavior of max-min fairness optimization algorithm with statistical \ac{CSI} for the alternating optimization algorithm. It shows that the algorithm converges in a reasonable number of iterations.

\section{CONCLUSION}
\label{conclusion}
Max-min fairness optimization is crucial to guaranteeing fair performance for all users in a \ac{SIM}-assisted multi-user system. In this paper, we proposed two novel max-min fairness algorithms based on instantaneous \ac{CSI} and statistical \ac{CSI} for \ac{SIM}-assisted multi-user \ac{MISO} systems. First, we propose a novel max-min fairness algorithm using alternating optimization technique to jointly optimize the power allocation and the wave-based beamforming based on instantaneous \ac{CSI}. To this end, we utilize \ac{GP} and gradient descent-ascent algorithms. Next, we proposed a novel algorithm to maximize the average minimum achievable rate based on statistical \ac{CSI}, focusing on a bound on the average achievable rate, since an exact average rate is analytically intractable. Subsequent simulation results demonstrate significant performance improvement due
to the proposed optimization algorithms. These results further exhibit higher minimum rate than equal power allocation and random phase shift assignments for instantaneous \ac{CSI}. Additionally, our simulations show that the proposed upper bound for the maximum achievable average minimum rate is tight for low \ac{SNR} regime for statistical \ac{CSI}. Thus, it is plausible to reduce computational overhead utilizing max-min fairness optimization in conjunction with the statistical \ac{CSI}, since it eliminates the need to optimize during every coherence interval. Notwithstanding the above observation, though computationally not efficient, per coherent frame max-min optimization with instantaneous CSI  gives better results.

\appendices

\section{Proof of Lemma I}\label{ap1}
\renewcommand{\theequation}{A-\arabic{equation}}
\setcounter{equation}{0}
The gradient of $\gamma_k$ with respect to $\theta_m^{(\ell)}$ can be expressed as 
\begin{equation}
    \frac{\partial f}{\partial {\theta_m^{(\ell)}}} = \sum_{k=1}^K {\lambda_k   \frac{\partial \gamma_k}{\partial {\theta_m^{(\ell)}}}}, 
\end{equation}
Subsequently, using the derivative of the ratio of two differentiable functions, i.e., the quotient rule, we can obtain $\frac{\partial \gamma_k}{\partial {\theta_m^{(\ell)}}}$ as
\begin{align}
    \frac{\partial \gamma_k}{\partial {\theta_m^{(\ell)}}}  &= \frac{p_k}{\sum_{\substack{j=1 \\ j \neq k}}^K|\mathbf{h}_k\herm \mathbf{G}_{\bm{\vartheta}}\mathbf{w}_j|^2p_j + \sigma^2_k} \frac{\partial |{\mathbf{h}_k\herm} \mathbf{G}_{\bm{\vartheta}}\mathbf{w}_{k}|^2}{\partial {\theta_m^{(\ell)}}} \notag \\
    &- \frac{|{\mathbf{h}_k\herm} \mathbf{G}_{\bm{\vartheta}}\mathbf{w}_{k}|^2p_k}{(\sum_{\substack{j=1 \\ j \neq k}}^K|\mathbf{h}_k\herm \mathbf{G}_{\bm{\vartheta}}\mathbf{w}_j|^2p_j + \sigma^2_k)^2}\sum_{\substack{j=1 \\ j \neq k}}^K p_j \frac{\partial |{\mathbf{h}_k\herm} \mathbf{G}_{\bm{\vartheta}}\mathbf{w}_{j}|^2}{\partial {\theta_m^{(\ell)}}},
\end{align}
where $\frac{\partial |{\mathbf{h}_k\herm} \mathbf{G}_{\bm{\vartheta}}\mathbf{w}_{j}|^2}{\partial {\theta_m^{(\ell)}}}$ can be expressed as 
\begin{align}
     \frac{\partial |{\mathbf{h}_k\herm} \mathbf{G}_{\bm{\vartheta}}\mathbf{w}_{j}|^2}{\partial {\theta_m^{(\ell)}}} &= \frac{\partial|\sum_{m=1}^M e^{\mathsf{j}\theta_m^{(\ell)}}\mathbf{h}_k\herm\mathbf{b}_m^{(\ell)}(\mathbf{a}^\ell_m)\herm\mathbf{w}_{j}|^2}{\partial{\theta^{(\ell)}_m}}  ,\notag \\
     &=     \frac{\partial\sum_{m=1}^Me^{\mathsf{j}\theta_m^{(\ell)}}\mathbf{z}_{m,l,k,j}\sum_{m=1}^Me^{-\mathsf{j}\theta_m^{(\ell)}}\mathbf{z}_{m,l,k,j}^*}{\partial{\theta^{(\ell)}_m}}  ,\notag \\     
     &=   \mathsf{j}e^{\mathsf{j}\theta_m^{(\ell)}}\mathbf{z}_{m,l,k,j}\sum_{m=1}^Me^{-\mathsf{j}\theta_m^{(\ell)}}\mathbf{z}_{m,l,k,j}^* \notag \\ &+ (-\mathsf{j})e^{-\mathsf{j}\theta_m^{(\ell)}}\mathbf{z}_{m,l,k,j}^*\sum_{m=1}^Me^{\mathsf{j}\theta_m^{(\ell)}}\mathbf{z}_{m,l,k,j}, \notag \\
     &\stackrel{\text{a}}{=}  2 \Re\left[ (-\mathsf{j})e^{-\mathsf{j}\theta_m^{(\ell)}}\mathbf{z}_{m,l,k,j}^*\sum_{m=1}^Me^{\mathsf{j}\theta_m^{(\ell)}}\mathbf{z}_{m,l,k,j}\right], \notag  \\
     &= 2\Im\left[\left(e^{\mathsf{j}\theta_m^{(\ell)}}\mathbf{h}_k\herm\mathbf{b}_m^{(\ell)}(\mathbf{a}^\ell_m)\herm\mathbf{w}_{j}\right)\herm({\mathbf{h}_k\herm} \mathbf{G}_{\bm{\vartheta}}\mathbf{w}_{j})\right],\notag \\
     &= \delta_{m,k,j}^{(\ell)},
\end{align}
where $\mathbf{z}_{m,l,k,j}=\mathbf{h}_k\herm\mathbf{b}_m^{(\ell)}(\mathbf{a}^\ell_m)\herm\mathbf{w}_{j} $. The step $\stackrel{\text{a}}{=}$ results from the fact that the gradient is real-valued. Substituting $\delta_{m,k,j}^{(\ell)}$ into the gradient expression and factoring out $\omega_k$ yields the final result, which completes the proof. \hfill $\blacksquare$ \vspace{1mm}

% \subsubsection{Proof of gradient in equation \ref{psi_update2}}
\section{Proof of Lemma II}\label{ap2}
\renewcommand{\theequation}{B-\arabic{equation}}
\setcounter{equation}{0}

It seems intractable to obtain gradient of $   \frac{1}{ \beta_k p_k \mathbf{w}^H_{k}\mathbf{G}_{\bm{\vartheta}}^H\mathbf{R}_{\text{RIS}} \mathbf{G}_{\bm{\vartheta}}\mathbf{w}_{k}} $. Hence, we eigen decompose the $\mathbf{R}_{\text{RIS}}$ as $ \mathbf{R}_{\text{RIS}} = \sum_{i=1}^M \rho_i \bm{\upsilon}_i\bm{\upsilon}_i\herm $ where $\rho_i$ and $\bm{\upsilon}_i$ are  eigenvalues and eigenvectors of $\mathbf{R}_{\text{RIS}}$, respectively. Thus, it can be written that
\begin{equation}   \mathbf{w}^H_{k}\mathbf{G}_{\bm{\vartheta}}^H\mathbf{R}_{\text{RIS}} \mathbf{G}_{\bm{\vartheta}}\mathbf{w}_{k} = \sum_{i=1}^M \rho_i|\bm{\upsilon}_i\herm\mathbf{G}_{\bm{\vartheta}}\mathbf{w}_{k}|^2 .
\end{equation} 
Subsequently, using the derivative of the reciprocal of a function, i.e, reciprocal rule, we can express $\frac{\partial \zeta}{\partial {\theta_m^{(\ell)}}}$ as
\begin{align}
    \frac{\partial \zeta}{\partial {\theta_m^{(\ell)}}} &= \sum_{k=1}^K    \frac{\partial }{\partial {\theta_m^{(\ell)}}} \left (\frac{1}{\beta_k p_k \mathbf{w}^H_{k}\mathbf{G}_{\bm{\vartheta}}^H\mathbf{R}_{\text{RIS}} \mathbf{G}_{\bm{\vartheta}}\mathbf{w}_{k}} \right), \notag \\    
    &= \sum_{k=1}^K    \frac{\partial }{\partial {\theta_m^{(\ell)}}} \left (\frac{1}{\beta_k p_k \sum_{i=1}^M \rho_i|\bm{\upsilon}_i\herm\mathbf{G}_{\bm{\vartheta}}\mathbf{w}_{k}|^2} \right), \notag \\
    &= \sum_{k=1}^K \frac{1}{\beta_k p_k} \frac{- \frac{\partial }{\partial {\theta_m^{(\ell)}}} \left (\sum_{i=1}^M \rho_i|\bm{\upsilon}_i\herm\mathbf{G}_{\bm{\vartheta}}\mathbf{w}_{k}|^2 \right)}{\left (\sum_{i=1}^M \rho_i|\bm{\upsilon}_i\herm\mathbf{G}_{\bm{\vartheta}}\mathbf{w}_{k}|^2 \right)^2}.  
\end{align}
Next, by recalling Appendix \ref{ap1}, we can express $ \frac{\partial }{\partial {\theta_m^{(\ell)}}} \left (\sum_{i=1}^M \rho_i|\bm{\upsilon}_i\herm\mathbf{G}_{\bm{\vartheta}}\mathbf{w}_{k}|^2 \right)$ as
\begin{align}
   &\frac{\partial }{\partial {\theta_m^{(\ell)}}} \left (\sum_{i=1}^M \rho_i|\bm{\upsilon}_i\herm\mathbf{G}_{\bm{\vartheta}}\mathbf{w}_{k}|^2 \right)= \sum_{i=1}^M \rho_i\frac{\partial }{\partial {\theta_m^{(\ell)}}} \left (|\bm{\upsilon}_i\herm\mathbf{G}_{\bm{\vartheta}}\mathbf{w}_{k}|^2 \right), \notag \\
   &\hspace{15mm}=  2\sum_{i=1}^M \rho_i \Im\left[ e^{-\mathsf{j}\theta_m^{(\ell)}}\mathbf{w}_k\herm\mathbf{a}^\ell_m(\mathbf{b}^\ell_m)\herm\bm{\upsilon}_i\bm{\upsilon}_i\herm \mathbf{G}_{\bm{\vartheta}}\mathbf{w}_k\right],\notag \\
   &\hspace{15mm}= c_{m,k}^{(\ell)}.
\end{align}

Summing over all eigenmodes $i = 1, \ldots, M$ and users $k = 1, \ldots, K$, weighted by $\rho_i$ and $\frac{1}{\beta_k p_k}$, yields the final expression for $\frac{\partial \zeta}{\partial \theta_m^{(\ell)}}$. This completes the proof. \hfill $\blacksquare$

\bibliographystyle{IEEEtran}
\bibliography{main}

% Generated by IEEEtran.bst, version: 1.14 (2015/08/26)
\begin{thebibliography}{10}
\providecommand{\url}[1]{#1}
\csname url@samestyle\endcsname
\providecommand{\newblock}{\relax}
\providecommand{\bibinfo}[2]{#2}
\providecommand{\BIBentrySTDinterwordspacing}{\spaceskip=0pt\relax}
\providecommand{\BIBentryALTinterwordstretchfactor}{4}
\providecommand{\BIBentryALTinterwordspacing}{\spaceskip=\fontdimen2\font plus
\BIBentryALTinterwordstretchfactor\fontdimen3\font minus \fontdimen4\font\relax}
\providecommand{\BIBforeignlanguage}[2]{{%
\expandafter\ifx\csname l@#1\endcsname\relax
\typeout{** WARNING: IEEEtran.bst: No hyphenation pattern has been}%
\typeout{** loaded for the language `#1'. Using the pattern for}%
\typeout{** the default language instead.}%
\else
\language=\csname l@#1\endcsname
\fi
#2}}
\providecommand{\BIBdecl}{\relax}
\BIBdecl

\bibitem{SIM_intro}
J.~An, M.~Di~Renzo, M.~Debbah, and C.~Yuen, ``Stacked intelligent metasurfaces for multiuser beamforming in the wave domain,'' in \emph{2023 IEEE Int. Conf. Commun. (ICC)}, 2023, pp. 2834--2839.

\bibitem{SIM_intro2}
J.~An \emph{et~al.}, ``Stacked intelligent metasurfaces for efficient holographic {MIMO} communications in {6G},'' \emph{IEEE J. Sel. Areas Commun.}, vol.~41, no.~8, pp. 2380--2396, 2023.

\bibitem{SIM_BD_RIS}
M.~Nerini and B.~Clerckx, ``Physically consistent modeling of stacked intelligent metasurfaces implemented with beyond diagonal {RIS},'' \emph{IEEE Commun. Lett.}, vol.~28, no.~7, pp. 1693--1697, 2024.

\bibitem{SIM_intro3}
J.~An \emph{et~al.}, ``Stacked intelligent metasurface-aided {MIMO} transceiver design,'' \emph{IEEE Wireless Commun.}, vol.~31, no.~4, pp. 123--131, 2024.

\bibitem{direnzo2024stateartstackedintelligent}
\BIBentryALTinterwordspacing
M.~D. Renzo, ``State-of-the-art on stacked intelligent metasurfaces: Communication, sensing and computing in the wave domain,'' 2024. [Online]. Available: \url{https://arxiv.org/abs/2411.19687}
\BIBentrySTDinterwordspacing

\bibitem{optimization_survey}
Y.-F. Liu \emph{et~al.}, ``A survey of recent advances in optimization methods for wireless communications,'' \emph{IEEE J. Sel. Areas Commun.}, vol.~42, no.~11, pp. 2992--3031, 2024.

\bibitem{optimization_wireless_intro}
X.~Lin, N.~Shroff, and R.~Srikant, ``A tutorial on cross-layer optimization in wireless networks,'' \emph{IEEE J. Sel. Areas Commun.}, vol.~24, no.~8, pp. 1452--1463, 2006.

\bibitem{optimization_wireless_intro2}
C.~W. Tan, ``Wireless network optimization by perron-frobenius theory,'' in \emph{2014 48th Annual Conference on Information Sciences and Systems (CISS)}, 2014, pp. 1--6.

\bibitem{power_opt_GP}
M.~Chiang, C.~W. Tan, D.~P. Palomar, D.~O'neill, and D.~Julian, ``Power control by geometric programming,'' \emph{IEEE Trans. Wireless Commun.}, vol.~6, no.~7, pp. 2640--2651, 2007.

\bibitem{BD_RIS_optimization}
A.~Sousa~de Sena, M.~Rasti, N.~Huda~Mahmood, and M.~Latva-aho, ``Beyond diagonal {RIS} for multi-band multi-cell {MIMO} networks: A practical frequency-dependent model and performance analysis,'' \emph{{IEEE} Trans. Wireless Commun.}, vol.~24, no.~1, pp. 749--766, 2025.

\bibitem{max_min_def}
B.~Radunovic and J.-Y. Le~Boudec, ``A unified framework for max-min and min-max fairness with applications,'' \emph{IEEE/ACM Trans. Netw.}, vol.~15, no.~5, pp. 1073--1083, 2007.

\bibitem{max_min_intro2}
D.~W.~H. Cai, T.~Q.~S. Quek, and C.~W. Tan, ``A unified analysis of max-min weighted {SINR} for {MIMO} downlink system,'' \emph{IEEE Trans. Signal Process.}, vol.~59, no.~8, pp. 3850--3862, 2011.

\bibitem{max_min_intro}
S.~He, Y.~Huang, L.~Yang, A.~Nallanathan, and P.~Liu, ``A multi-cell beamforming design by uplink-downlink max-min {SINR} duality,'' \emph{IEEE Trans. Wireless Commun.}, vol.~11, no.~8, pp. 2858--2867, 2012.

\bibitem{SIM_optimization_statistical_CSI}
A.~Papazafeiropoulos, P.~Kourtessis, S.~Chatzinotas, D.~I. Kaklamani, and I.~S. Venieris, ``Achievable rate optimization for large stacked intelligent metasurfaces based on statistical {CSI},'' \emph{IEEE Wireless Commun. Lett.}, vol.~13, no.~9, pp. 2337--2341, 2024.

\bibitem{an2023stackedintelligentmetasurfacesmultiuser}
\BIBentryALTinterwordspacing
J.~An, M.~D. Renzo, M.~Debbah, H.~V. Poor, and C.~Yuen, ``Stacked intelligent metasurfaces for multiuser downlink beamforming in the wave domain,'' 2023. [Online]. Available: \url{https://arxiv.org/abs/2309.02687}
\BIBentrySTDinterwordspacing

\bibitem{SIM_optimization_holographic_MIMO}
A.~Papazafeiropoulos, J.~An, P.~Kourtessis, T.~Ratnarajah, and S.~Chatzinotas, ``Achievable rate optimization for stacked intelligent metasurface-assisted holographic {MIMO} communications,'' \emph{IEEE Trans. Wireless Commun.}, vol.~23, no.~10, pp. 13\,173--13\,186, 2024.

\bibitem{SIM_fully_analog_beamforming}
Z.~Li, J.~An, and C.~Yuen, ``Stacked intelligent metasurfaces for fully-analog wideband beamforming design,'' in \emph{2024 IEEE VTS Asia Pacific Wireless Communications Symposium (APWCS)}, 2024, pp. 1--5.

\bibitem{SIM_optimization_DRL}
\BIBentryALTinterwordspacing
H.~Liu, J.~An, G.~C. Alexandropoulos, D.~W.~K. Ng, C.~Yuen, and L.~Gan, ``Multi-user {MISO} with stacked intelligent metasurfaces: a {DRL}-based sum-rate optimization approach,'' 2024. [Online]. Available: \url{https://arxiv.org/abs/2408.04837}
\BIBentrySTDinterwordspacing

\bibitem{SIM_beamforming_DRL}
H.~Liu, J.~An, D.~W.~K. Ng, G.~C. Alexandropoulos, and L.~Gan, ``{DRL}-based orchestration of multi-user {MISO} systems with stacked intelligent metasurfaces,'' in \emph{ICC 2024 - IEEE International Conference on Communications}, 2024, pp. 4991--4996.

\bibitem{hybrid_SIM}
D.~Darsena, F.~Verde, I.~Iudice, and V.~Galdi, ``Design of stacked intelligent metasurfaces with reconfigurable amplitude and phase for multiuser downlink beamforming,'' \emph{IEEE Open Journal of the Communications Society}, vol.~6, pp. 531--550, 2025.

\bibitem{shi2024harnessing}
E.~Shi, J.~Zhang, Y.~Zhu, J.~An, C.~Yuen, and B.~Ai, ``Harnessing stacked intelligent metasurface for enhanced cell-free massive {MIMO} systems: A low-power and cost approach,'' \emph{arXiv preprint arXiv:2409.12851}, 2024.

\bibitem{SIM_ISAC}
S.~Li, F.~Zhang, T.~Mao, R.~Na, Z.~Wang, and G.~K. Karagiannidis, ``Transmit beamforming design for {ISAC} with stacked intelligent metasurfaces,'' \emph{IEEE Trans. Veh. Technol.}, pp. 1--6, 2024.

\bibitem{max_min_group_RIS}
H.~Cheng, J.~Huang, G.~Li, F.~Wang, W.~Hao, and S.~Yang, ``Max -min rate optimization for group- transmissive {RIS}- based transmitter architectures,'' in \emph{2024 IEEE 99th Vehicular Technology Conference (VTC2024-Spring)}, 2024, pp. 1--5.

\bibitem{max_min_RIS}
H.~Xie, J.~Xu, and Y.-F. Liu, ``Max-min fairness in {IRS}-aided multi-cell {MISO} systems with joint transmit and reflective beamforming,'' \emph{IEEE Trans. Wireless Commun.}, vol.~20, no.~2, pp. 1379--1393, 2021.

\bibitem{Asymptotic_Max-Min_SINR_RIS}
Q.-U.-A. Nadeem, A.~Kammoun, A.~Chaaban, M.~Debbah, and M.-S. Alouini, ``Asymptotic max-min {SINR} analysis of reconfigurable intelligent surface assisted {MISO} systems,'' \emph{IEEE Trans. Wireless Commun.}, vol.~19, no.~12, pp. 7748--7764, 2020.

\bibitem{max_min_RIS_NOMA}
L.~Cantos, M.~Awais, and Y.~H. Kim, ``Max-min rate optimization for uplink {IRS-NOMA} with receive beamforming,'' \emph{IEEE Wireless Commun. Lett.}, vol.~11, no.~12, pp. 2512--2516, 2022.

\bibitem{max_min_RIS_cell_free2}
M.~Bashar, K.~Cumanan, A.~G. Burr, P.~Xiao, and M.~Di~Renzo, ``On the performance of reconfigurable intelligent surface-aided cell-free massive {MIMO} uplink,'' in \emph{GLOBECOM 2020 - 2020 IEEE Global Communications Conference}, 2020, pp. 1--6.

\bibitem{Max_min_RIS_green}
A.~Subhash, A.~Kammoun, A.~Elzanaty, S.~Kalyani, Y.~H. Al-Badarneh, and M.-S. Alouini, ``Max-min {SINR} optimization for {RIS}-aided uplink communications with green constraints,'' \emph{IEEE Wireless Commun. Lett.}, vol.~12, no.~6, pp. 942--946, 2023.

\bibitem{max_min_RIX_convex_hull_relaxation}
W.~Lai, Z.~Wu, Y.~Feng, K.~Shen, and Y.-F. Liu, ``An efficient convex-hull relaxation based algorithm for multi-user discrete passive beamforming,'' \emph{IEEE Signal Process. Lett.}, vol.~31, pp. 2275--2279, 2024.

\bibitem{max_min_RIS_GDA}
G.~Yan, L.~Zhu, and R.~Zhang, ``Passive refleccommuntion optimization for {IRS}-aided multicast beamforming with discrete phase shifts,'' \emph{IEEE Wireless Commun. Lett.}, vol.~12, no.~8, pp. 1424--1428, 2023.

\bibitem{max_min_RIS_cell_Free}
Q.-U.-A. Nadeem, A.~Zappone, and A.~Chaaban, ``Achievable rate analysis and max-min {SINR} optimization in intelligent reflecting surface assisted cell-free {MIMO} uplink,'' \emph{IEEE Open J. Commun. Soc.}, vol.~3, pp. 1295--1322, 2022.

\bibitem{max_min_star_RIS}
A.~Papazafeiropoulos, P.~Kourtessis, and S.~Chatzinotas, ``Max-min {SINR} analysis of {STAR-RIS} assisted massive {MIMO} systems with hardware impairments,'' \emph{IEEE Trans. Wireless Commun.}, vol.~23, no.~5, pp. 4255--4268, 2024.

\bibitem{S_CSI_RIS_optimization}
D.~Gunasinghe and G.~Amarasuriya, ``Statistical {CSI} based phase-shift and transmit power optimization for {RIS}-aided massive {MIMO},'' in \emph{GLOBECOM 2022 - 2022 IEEE Global Communications Conference}, 2022, pp. 4007--4012.

\bibitem{RIS_statisticalCSI}
Z.~Yu, Y.~Han, M.~Matthaiou, X.~Li, and S.~Jin, ``Statistical {CSI}-based design for {RIS}-assisted communication systems,'' \emph{IEEE Wireless Commun. Lett.}, vol.~11, no.~10, pp. 2115--2119, 2022.

\bibitem{RIS_statisticalCSI2}
K.~Zhi, C.~Pan, H.~Ren, and K.~Wang, ``Statistical {CSI}-based design for reconfigurable intelligent surface-aided massive {MIMO} systems with direct links,'' \emph{IEEE Wireless Commun. Lett.}, vol.~10, no.~5, pp. 1128--1132, 2021.

\bibitem{RIS_statisticalCSI_DRL}
M.~Eskandari, H.~Zhu, A.~Shojaeifard, and J.~Wang, ``Statistical {CSI}-based beamforming for {RIS}-aided multiuser {MISO} systems via deep reinforcement learning,'' \emph{IEEE Wireless Commun. Lett.}, vol.~13, no.~2, pp. 570--574, 2024.

\bibitem{doi:10.1126/science.aat8084}
\BIBentryALTinterwordspacing
X.~Lin \emph{et~al.}, ``All-optical machine learning using diffractive deep neural networks,'' \emph{Science}, vol. 361, no. 6406, pp. 1004--1008, 2018. [Online]. Available: \url{https://www.science.org/doi/abs/10.1126/science.aat8084}
\BIBentrySTDinterwordspacing

\bibitem{boyd2004convex}
S.~Boyd and L.~Vandenberghe, \emph{Convex Optimization}.\hskip 1em plus 0.5em minus 0.4em\relax Cambridge, UK: Cambridge University Press, 2004.

\bibitem{lin2025twotimeGDA}
T.~Lin, C.~Jin, and M.~I. Jordan, ``Two-timescale gradient descent ascent algorithms for nonconvex minimax optimization,'' \emph{Journal of Machine Learning Research}, vol.~26, no.~11, pp. 1--45, 2025.

\bibitem{uatf_bound}
L.~Sanguinetti, E.~Björnson, and J.~Hoydis, ``Toward massive {MIMO} 2.0: Understanding spatial correlation, interference suppression, and pilot contamination,'' \emph{IEEE Trans. Commun.}, vol.~68, no.~1, pp. 232--257, 2020.

\bibitem{FI_intro1}
M.~Dianati, X.~Shen, and S.~Naik, ``A new fairness index for radio resource allocation in wireless networks,'' in \emph{IEEE Wireless Communications and Networking Conference, 2005}, vol.~2, 2005, pp. 712--717 Vol. 2.

\bibitem{FI_intro2}
R.~K. Jain, D.-M.~W. Chiu, W.~R. Hawe \emph{et~al.}, ``A quantitative measure of fairness and discrimination,'' \emph{Eastern Research Laboratory, Digital Equipment Corporation, Hudson, MA}, vol.~21, no.~1, 1984.

\bibitem{FI_intro3}
\BIBentryALTinterwordspacing
U.~U. Khan, N.~Dilshad, M.~H. Rehmani, and T.~Umer, ``Fairness in cognitive radio networks: Models, measurement methods, applications, and future research directions,'' \emph{Journal of Network and Computer Applications}, vol.~73, pp. 12--26, 2016. [Online]. Available: \url{https://www.sciencedirect.com/science/article/pii/S1084804516301527}
\BIBentrySTDinterwordspacing

\bibitem{FI_intro4}
\BIBentryALTinterwordspacing
Y.~Cui, P.~Liu, Y.~Zhou, and W.~Duan, ``Energy-efficient resource allocation for downlink non-orthogonal multiple access systems,'' \emph{Applied Sciences}, vol.~12, no.~19, 2022. [Online]. Available: \url{https://www.mdpi.com/2076-3417/12/19/9740}
\BIBentrySTDinterwordspacing

\end{thebibliography}
\end{document}